\documentclass[usenatbib]{mnras}
\usepackage{graphicx}

\title[SN2013L]{Optical and IR observations of SN 2013L, a Type IIn Supernova surrounded by asymmetric CSM }

\author[Andrews et al.]{Jennifer E. Andrews$^1$\thanks{Email:
    jandrews@as.arizona.edu}, Nathan Smith$^1$, Curtis McCully$^{2,3}$, Ori D. Fox$^4$, S. Valenti$^5$,
     \newauthor D.A. Howell$^{2,3}$\\
     $^1$Steward Observatory, University of Arizona, 933 North Cherry Avenue, Tucson,AZ 85721, USA\\
     $^2$Las Cumbres Observatory, 6740 Cortona Dr Suite 102, Goleta, CA 93117-5575, USA\\
     $^3$ Department of Physics, University of California, Santa Barbara, CA 93106-9530, USA\\
     $^4$Space Telescope Science Institute, 3700 San Martin Drive, Baltimore, MD 21218, USA\\
     $^5$Department of Physics, University of California, Davis, 1 Shields Ave, Davis, CA 95616, USA\\
      }

\date{Accepted XXX. Received YYY; in original form ZZZ}

\pubyear{2017}

\begin{document}
\label{firstpage}
\pagerange{\pageref{firstpage}--\pageref{lastpage}}
\maketitle

\begin{abstract}

We present optical and NIR photometry and spectroscopy of SN 2013L for the first four years post-explosion. SN 2013L was a moderately luminous  (M$_{r}$ = -19.0) Type IIn supernova (SN) that showed signs of strong shock interaction with the circumstellar medium (CSM).  The CSM interaction was equal to or stronger to SN 1988Z for the first 200 days and is observed at all epochs after explosion. Optical spectra revealed multi-component hydrogen lines appearing by day 33 and persisting and slowly evolving over the next few years.  By day 1509 the H$\alpha$ emission was still strong and exhibiting multiple peaks, hinting that the CSM was in a disc or torus  around the SN.  SN 2013L is part of a growing subset of SNe IIn that shows both strong CSM interaction signatures and the underlying broad lines from the SN ejecta photosphere. The presence of a blue H$\alpha$ emission bump and a lack of a red peak does not appear to be due to dust obscuration since an identical profile is seen in Pa$\beta$.  Instead this suggests a high concentration of material on the near-side of the SN or a disc inclination of roughly edge-on and hints that SN 2013L was part of a massive interactive binary system. Narrow H$\alpha$ P-Cygni lines that persist through the entirety of the observations measure a progenitor outflow speed of 80--130 km s$^{-1}$, speeds normally associated  with extreme red supergiants, yellow hypergiants, or luminous blue variable winds. This progenitor scenario is also consistent with an inferred progenitor mass-loss rate of  0.3 - 8.0 $\times$ 10$^{-3}$ M$_{\sun}$ yr$^{-1}$.
\end{abstract}


\section{INTRODUCTION}
The Type II class of core collapse supernovae (CCSNe) come from massive stars ($>$ 8 M$_{\sun}$) that still contain at least some portion of their hydrogen envelope. Approximately 8-9$\%$ of CCSNe show the  presence of narrow ($\sim$ 100 km s$^{-1}$) hydrogen emission lines in their spectra \citep{2011MNRAS.412.1522S}, due to the  photoionization or shock heating of the surrounding, slow moving circumstellar medium (CSM). These CCSNe are classified as Type IIn  \citep{1990MNRAS.244..269S}, where the 'n' stands for narrower lines than seen in normal Type II SNe.

Depending on the extent and density of the CSM, some SNe IIn show narrow lines only fleetingly \citep{2015MNRAS.449.1876S,2014Natur.509..471G} and others can exhibit additional broad ($\sim$ 10000 km s$^{-1}$) and intermediate ($\sim$ 1000 km s$^{-1}$) hydrogen emission lines for their full duration.  The broad component traces emission from the free expansion of the SN ejecta, while the intermediate components are formed in the post-shock region between the forward and reverse shocks created as the ejecta moves through the CSM.   This shock interaction transforms the kinetic energy of the ejecta into radiative energy, adding additional luminosity to the SN light curve.  For some objects, this interaction can keep the late-time light curve bright for months or years after explosion \citep{1999MNRAS.309..343A,2002MNRAS.333...27P,2012ApJ...756..173S,2013AJ....146....2F,2015MNRAS.454.4366F,2017MNRAS.466.3021S}. See  \citet{2016arXiv161202006S} for a review of CSM interacting SNe.

While the presence of narrow emission lines is the unifying factor for these SNe, they are far from homogeneous. They can span absolute magnitudes between -22 and -17,  have progenitor mass-loss rates between 10$^{-4}$ -- 1 M$_{\sun}$ yr$^{-1}$, and wind velocities between 40 - 1000 km s$^{-1}$ \citep{2012ApJ...744...10K,2007ApJ...666.1116S, 2008ApJ...686..467S,1998MNRAS.294..448B,2013A&A...555A..10T}.  These differences arise due to a variety of different progenitors and mass-loss histories (see \citet{2016arXiv161202006S} for a full review).  The progenitors for IIn showing signs of slower wind velocities (40 -100 km s$^{-1}$) and smaller mass-loss rates (10$^{-4}$ -- 10$^{-3}$ M$_{\sun}$ yr$^{-1}$) are likely  extreme red supergiant (RSG) or yellow hypergiant (YHG) progenitors \citep{2014ARA&A..52..487S}. For instance the IIn SNe SN 2005ip \citep{2017MNRAS.466.3021S}, PTF11iqb \citep{2015MNRAS.449.1876S}, SN 2013fc \citep{2016MNRAS.456..323K}, and SN 1998S \citep{2012MNRAS.424.2659M} likely had RSG progenitors. Luminous blue variables (LBVs), traditionally not considered as a terminal point for massive star evolution, have been cited as progenitors of IIn  that show signs of moderately fast winds ($\sim$ 100--600 km s$^{-1}$) and high mass-loss rates \citep{2013MNRAS.431.2599M, 2009Natur.458..865G, 2011MNRAS.415..773S,2007ApJ...666.1116S,2008ApJ...686..467S,2010AJ....139.1451S}, but see \citet{2017ApJ...836..222F} for an alternative progenitor scenario of SN 2010jl.  These progenitors have masses $>$ 30 M$_{\sun}$, and sometime show signs of pre-supernova outbursts \citep{2013MNRAS.430.1801M,2013ApJ...767....1P,2016ApJ...824....6O}, although not always \citep{2015MNRAS.450..246B}. Interestingly, it appears that  SNe IIn are not strongly correlated with H$\alpha$ regions   \citep{2016arXiv160806097K,2014MNRAS.441.2230H}. This may hint that RSGs are the likely progenitors for for all but the most luminous Type IIn \citep{2009AJ....137.3558S}.

To date, no definitive progenitor scenario seems to exist for these interacting SNe, and it is likely that they can arise from a wide variety of progenitor masses. Additionally, it is important to remember that up to 75$\%$ of massive stars are in binaries with separations close enough for interaction \citep{2012Sci...337..444S, 2012ApJ...751....4K, 2014ApJ...782....7D}, therefore we should not neglect binary interaction as a potential channel for IIn production.  Instabilities in a single massive progenitor and/or the presence of a companion can lead to sustained and often asymmetric mass-loss \citep{2014ApJ...785...82S,2012MNRAS.423L..92Q}. Whether from a single  star or a binary system, when significant mass-loss events occur prior to core collapse, the result is narrow lines soon after explosion.

\begin{figure}
\includegraphics[width=3.3in]{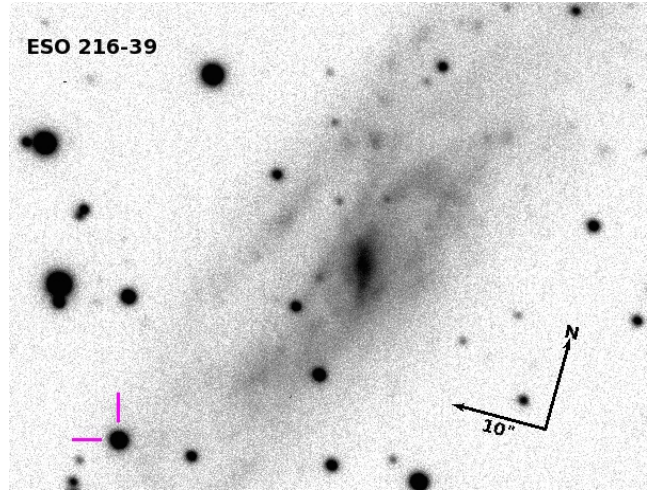}
\caption{Magellan/IMACS R-band image of SN2013L taken on 2013 April 16 (day 84).  The SN is indicated by the magenta cross hairs.}
\label{fig:finder}
\end{figure}

SN2013L ($\rmn{RA}(2000)=11^{\rmn{h}} 45^{\rmn{m}} 29\fs55$, $\rmn{Dec.}~(2000)=-50\degr 35\arcmin 53\farcs 1$) was discovered in a southeastern spiral arm of ESO 216-39 on 2013 January 22, and classified as a young Type IIn supernova shortly thereafter based on the presence of narrow Balmer emission lines  \citep[Figure 1]{2013CBET.3392....1M}. The exact explosion date, while likely within a few days of the discovery, is unknown so in this paper all epochs will be relative to the date of discovery.  Using emission lines in the HII region associated with the SN we measure a redshift z=0.01738. Assuming H$_{0}$ = 71.9 km s$^{-1}$ Mpc$^{-1}$ \citep{2017MNRAS.465.4914B}, this gives an uncorrected rotational distance of 72 Mpc.  NED\footnote{The NASA/IPAC Extragalactic Database (NED) is operated by the Jet Propulsion Laboratory, California Institute of Technology, under contract with the National Aeronautics and Space Administration.} lists a Galactocentric  distance of 67 Mpc, a distance of 73.4 corrected for the motion with respect to the 3K CMB, and a distance corrected for the Great Attractor and Virgo of 80.5 Mpc. Therefore we adopt a distance of 72$^{+8.5}_{-5}$ Mpc throughout the paper.  If the unfiltered discovery magnitude of 15.6 was peak brightness and if we adopt the Milky Way line-of-sight reddening of E(B-V) = 0.11 mag \citep{1998ApJ...500..525S}, this gives an estimated absolute magnitude at peak of -19.0$\pm{0.2}$.  Here we present the optical photometric and spectroscopic evolution of SN 2013L from 5 to 1509 days after discovery.  The data and their reduction will be discussed in Section 2, the photometric and spectroscopic evolution in Sections 3 and 4, the implications for the progenitor in Section 5, and finally a summary of conclusions will be presented in Section 6.

\section{OBSERVATIONS}
 \subsection{Imaging}
 
\begin{figure*}
\includegraphics[width=6.0in]{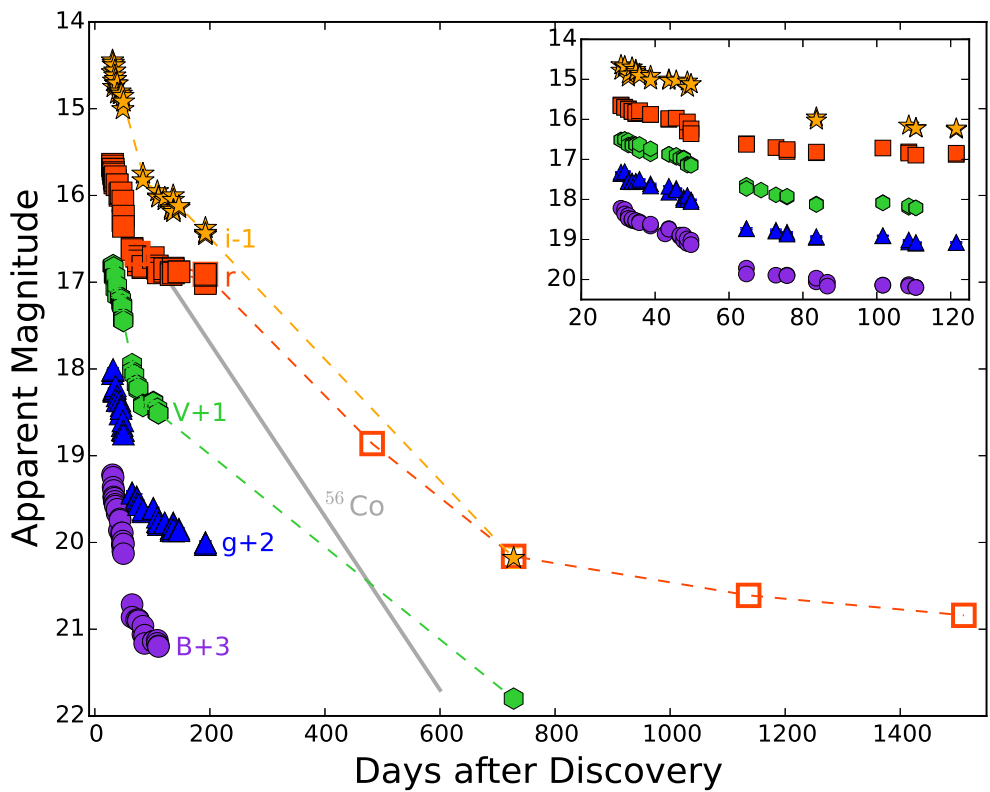}
\caption{Lightcurve of SN2013L in multiple bands.  Filters have been shifted for clarity and are indicated on the plot.  The inset shows a blow-up of the first 120 days, and the grey solid line indicates the radioactive decay of $^{56}$Co. R-band Magellan/IMACS photometry is indicated by the open squares. The photometry have not been corrected for extinction and come directly from the values listed in Table 1.}
\label{fig:fulllc}
\end{figure*}

Starting on day 30, optical monitoring in g$^{\prime}$r$^{\prime}$i$^{\prime}$BV filters was conducted using the Las Cumbres Observatory network\footnote{http://lcogt.net/}  \citep[LCO]{2013PASP..125.1031B}.  These observations constitute our early light curve data, seen in Figure \ref{fig:fulllc}.  The images were reduced using the {\tt lcogtsnpipe} pipeline \citep{2016MNRAS.459.3939V}, and photometric calibration was done using APASS\footnote{https://www.aavso.org/apass} standards. 

We also obtained 20 second R band exposures with Magellan/IMACS \citep{2011PASP..123..288D} on days 84, 482, 1133, and 1509 after explosion.  On day 728, 60 second exposures in each VRi$^{\prime}$ filter were taken.  Standard reduction techniques with {\tt IRAF} \footnote{IRAF, the Image Reduction and Analysis Facility is distributed by the National Optical Astronomy Observatory, which is operated by the Association of Universities for Research in Astronomy (AURA) under cooperative agreement with the National Science Foundation (NSF)} were used, and absolute calibration was performed using APASS standards and aperture photometry. Transformation between r$^{\prime}$ and R for the standards was carried out using \citet{2005AJ....130..873J}, and tertiary standards were created across all IMACS images to create a robust zero point. The photometry is summarized in Table \ref{tab:photometry}.

\begin{figure}
\includegraphics[width=3.3in]{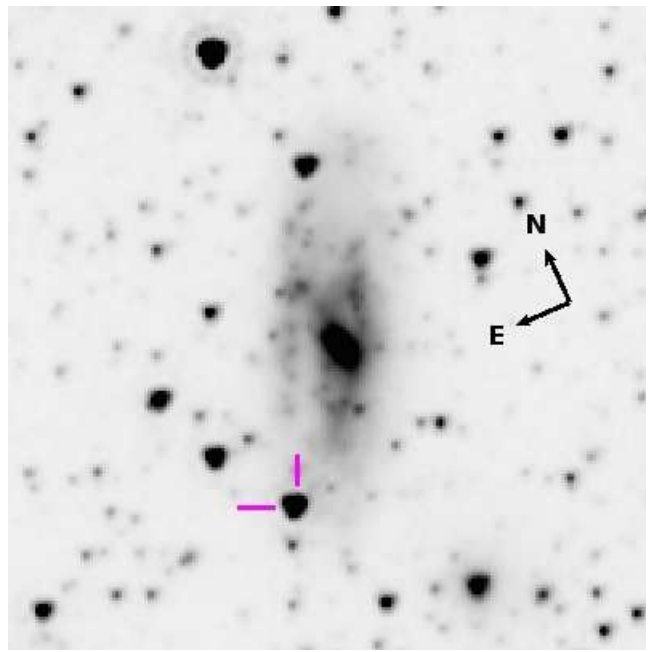}
\caption{Spitzer 4.5$\mu$m image of SN2013L on day 850.  The SN is indicated by magenta cross hairs.}
\label{fig:spitzimage}
\end{figure}

Spitzer IRAC (3.6 and 4.5 $\mu$m) images on days 630 and 850 were retrieved from the Spitzer Heritage Archive as part of an ongoing program by PI Fox (PID 10139,11053). Aperture photometry was performed on the pipeline reduced \textit{pbcd} images, with the appropriate aperture corrections.  A summary of dates and fluxes are shown in Table \ref{tab:spit}, and an example of the images can be seen in Figure \ref{fig:spitzimage}.

\begin{table*}
\caption{Optical Photometry of SN~2013L obtained with LCO and Magellan/IMACS*}
\begin{tabular}{@{}lccccc}\hline\hline
JD&g&r&i&B&V\\ \hline
2456344.75	&	16.0559	$\pm$	0.0154	&	15.6389	$\pm$	0.0143	&	15.5815	$\pm$	0.0171	&	16.2195	$\pm$	0.0145	&	15.8145	$\pm$	0.0171	\\
2456344.76	&	16.0188	$\pm$	0.014	&	15.6595	$\pm$	0.0249	&	15.4453	$\pm$	0.0134	&		$\cdot\cdot\cdot$		&	15.8149	$\pm$	0.0163	\\
2456345.76	&	16.0206	$\pm$	0.0182	&	15.6941	$\pm$	0.0141	&	15.4923	$\pm$	0.0133	&	16.2489	$\pm$	0.0169	&	15.8051	$\pm$	0.0156	\\
2456345.76	&	16.0042	$\pm$	0.0184	&	15.6985	$\pm$	0.0156	&	15.4714	$\pm$	0.015	&	16.3627	$\pm$	0.0188	&	15.8028	$\pm$	0.019	\\
2456346.76	&	16.2382	$\pm$	0.0173	&	15.7351	$\pm$	0.0185	&	15.7514	$\pm$	0.0222	&	16.4003	$\pm$	0.0319	&	15.8293	$\pm$	0.0263	\\
2456346.76	&	16.2556	$\pm$	0.0128	&	15.7521	$\pm$	0.0169	&	15.6777	$\pm$	0.0231	&	16.4748	$\pm$	0.0229	&	15.9417	$\pm$	0.0198	\\
2456347.76	&	16.2368	$\pm$	0.0161	&	15.8027	$\pm$	0.0173	&	15.5017	$\pm$	0.0136	&	16.5229	$\pm$	0.0166	&	15.9215	$\pm$	0.0151	\\
2456347.76	&	16.2586	$\pm$	0.0164	&	15.8003	$\pm$	0.0171	&		$\cdot\cdot\cdot$		&	16.4904	$\pm$	0.0163	&	15.952	$\pm$	0.0205	\\
2456348.76	&	16.2554	$\pm$	0.0144	&	15.8567	$\pm$	0.0154	&	15.5615	$\pm$	0.02	&	16.5445	$\pm$	0.019	&	15.9284	$\pm$	0.0168	\\
2456348.76	&	16.2462	$\pm$	0.0128	&	15.824	$\pm$	0.0173	&	15.619	$\pm$	0.0154	&	16.5261	$\pm$	0.0193	&	15.933	$\pm$	0.0159	\\
2456349.76	&	16.2256	$\pm$	0.0175	&	15.8268	$\pm$	0.0162	&	15.6599	$\pm$	0.02	&	16.56	$\pm$	0.0241	&	16.0859	$\pm$	0.0144	\\
2456349.76	&	16.1953	$\pm$	0.016	&	15.7834	$\pm$	0.0211	&	15.7098	$\pm$	0.0209	&	16.5809	$\pm$	0.0216	&	15.9249	$\pm$	0.0152	\\
2456352.76	&	16.3108	$\pm$	0.0174	&	15.8777	$\pm$	0.0145	&	15.7102	$\pm$	0.0163	&	16.6634	$\pm$	0.013	&	16.162	$\pm$	0.0129	\\
2456352.76	&	16.3647	$\pm$	0.0145	&	15.8758	$\pm$	0.0115	&	15.8261	$\pm$	0.02	&	16.6168	$\pm$	0.0131	&	16.035	$\pm$	0.0129	\\
2456356.71	&		$\cdot\cdot\cdot$		&		$\cdot\cdot\cdot$		&		$\cdot\cdot\cdot$		&	16.8604	$\pm$	0.0281	&	$\cdot\cdot\cdot$			\\
2456357.72	&	16.5205	$\pm$	0.0108	&	16.0003	$\pm$	0.0116	&	15.8035	$\pm$	0.0116	&	16.728	$\pm$	0.0171	&	16.1711	$\pm$	0.0158	\\
2456357.72	&	16.3707	$\pm$	0.0115	&	15.9794	$\pm$	0.0108	&	15.8334	$\pm$	0.0107	&	16.7437	$\pm$	0.0174	&	16.1724	$\pm$	0.0121	\\
2456359.72	&	16.4253	$\pm$	0.0122	&	15.9693	$\pm$	0.0196	&	15.8283	$\pm$	0.0512	&	16.9236	$\pm$		&	16.2479	$\pm$	0.0805	\\
2456359.72	&	16.4569	$\pm$	0.0163	&		$\cdot\cdot\cdot$		&		$\cdot\cdot\cdot$		&		$\cdot\cdot\cdot$		&	16.1982	$\pm$	0.0359	\\
2456360.71	&		$\cdot\cdot\cdot$		&		$\cdot\cdot\cdot$		&	$\cdot\cdot\cdot$			&	16.9236	$\pm$	0.0167	&	16.2689	$\pm$	0.0214	\\
2456360.71	&		$\cdot\cdot\cdot$		&		$\cdot\cdot\cdot$		&	$\cdot\cdot\cdot$			&	16.8899	$\pm$	0.0216	&	16.261	$\pm$	0.0177	\\
2456361.72	&	16.6534	$\pm$	0.0268	&	$\cdot\cdot\cdot$			&			$\cdot\cdot\cdot$	&	17.0263	$\pm$	0.0158	&	16.2801	$\pm$	0.0194	\\
2456361.73	&	16.7027	$\pm$	0.0214	&	$\cdot\cdot\cdot$			&		$\cdot\cdot\cdot$		&	16.8886	$\pm$	0.0181	&	$\cdot\cdot\cdot$			\\
2456362.72	&	16.7052	$\pm$	0.0153	&	16.2899	$\pm$	0.0136	&	15.8535	$\pm$	0.0102	&	17.0989	$\pm$	0.018	&	16.4172	$\pm$	0.0153	\\
2456362.72	&	16.5996	$\pm$	0.0156	&	16.0636	$\pm$	0.0163	&	16.0068	$\pm$	0.0248	&	16.9833	$\pm$	0.0156	&	16.4432	$\pm$	0.0145	\\
2456363.72	&	16.7105	$\pm$	0.0177	&	16.2372	$\pm$	0.0196	&	15.919	$\pm$	0.03	&	17.0173	$\pm$	0.025	&	16.4207	$\pm$	0.0225	\\
2456363.72	&	16.7522	$\pm$	0.0254	&	16.3574	$\pm$	0.0241	&		$\cdot\cdot\cdot$		&	17.1292	$\pm$	0.0229	&	16.449	$\pm$	0.0215	\\
2456378.72	&	17.4258	$\pm$	0.021	&	16.6082	$\pm$	0.0245	&	$\cdot\cdot\cdot$			&	17.716	$\pm$	0.0311	&	16.9443	$\pm$	0.0287	\\
2456378.72	&	17.4339	$\pm$	0.0278	&	16.6235	$\pm$	0.0219	&	$\cdot\cdot\cdot$			&	17.8611	$\pm$	0.0371	&	17.0272	$\pm$	0.034	\\
2456382.65	&			$\cdot\cdot\cdot$	&			$\cdot\cdot\cdot$	&	$\cdot\cdot\cdot$			&	$\cdot\cdot\cdot$			&	17.0673	$\pm$	0.0322	\\
2456386.70	&	17.471	$\pm$	0.0305	&	16.7031	$\pm$	0.0296	&		$\cdot\cdot\cdot$		&	17.8922	$\pm$	0.0193	&	17.1843	$\pm$	0.0172	\\
2456386.70	&	17.4914	$\pm$	0.0108	&	$\cdot\cdot\cdot$			&	$\cdot\cdot\cdot$			&	$\cdot\cdot\cdot$			&	$\cdot\cdot\cdot$			\\
2456388.50*	&			$\cdot\cdot\cdot$	&	16.65	$\pm$	0.1	&	$\cdot\cdot\cdot$			&		$\cdot\cdot\cdot$		&		$\cdot\cdot\cdot$		\\
2456389.70	&	17.5188	$\pm$	0.0142	&	16.8021	$\pm$	0.0116	&	$\cdot\cdot\cdot$			&	17.8898	$\pm$	0.0169	&	17.2418	$\pm$	0.0155	\\
2456389.70	&	17.5657	$\pm$	0.0151	&	16.7536	$\pm$	0.0124	&	$\cdot\cdot\cdot$			&	17.9031	$\pm$	0.0154	&	17.2194	$\pm$	0.0199	\\
2456397.63	&	17.6481	$\pm$	0.0207	&	16.8336	$\pm$	0.0255	&	16.7504	$\pm$	0.0245	&	18.0585	$\pm$	0.0269	&	17.4132	$\pm$	0.0388	\\
2456397.63	&	17.6267	$\pm$	0.022	&	16.8172	$\pm$	0.0281	&	16.8276	$\pm$	0.0158	&	17.9653	$\pm$	0.0198	&	17.4284	$\pm$	0.0368	\\
2456400.62	&			$\cdot\cdot\cdot$	&		$\cdot\cdot\cdot$		&		$\cdot\cdot\cdot$		&	18.068	$\pm$	0.0254	&		$\cdot\cdot\cdot$		\\
2456400.62	&			$\cdot\cdot\cdot$	&		$\cdot\cdot\cdot$		&		$\cdot\cdot\cdot$		&	18.1644	$\pm$	0.0285	&	$\cdot\cdot\cdot$			\\
2456415.64	&	17.6046	$\pm$	0.0115	&	16.7153	$\pm$	0.0115	&		$\cdot\cdot\cdot$		&	18.1384	$\pm$	0.0156	&	17.3786	$\pm$	0.0165	\\
2456415.64	&	17.6094	$\pm$	0.0117	&			$\cdot\cdot\cdot$	&		$\cdot\cdot\cdot$		&	18.1416	$\pm$	0.0191	&	17.387	$\pm$	0.0148	\\
2456422.65	&	17.7628	$\pm$	0.0122	&	16.8116	$\pm$	0.0101	&	16.9529	$\pm$	0.0147	&	18.1274	$\pm$	0.0188	&	17.4978	$\pm$	0.0148	\\
2456422.65	&	17.7101	$\pm$	0.0104	&	16.841	$\pm$	0.01		&		$\cdot\cdot\cdot$		&	18.1631	$\pm$	0.0172	&	17.4575	$\pm$	0.0156	\\
2456424.58	&	17.7955	$\pm$	0.0154	&	16.8936	$\pm$	0.0147	&	17.0141	$\pm$	0.0214	&	18.1917	$\pm$	0.0147	&	17.5109	$\pm$	0.0183	\\
2456424.58	&	17.7745	$\pm$	0.0124	&		$\cdot\cdot\cdot$		&	17.0255	$\pm$	0.0205	&	18.1998	$\pm$	0.0149	&	17.5116	$\pm$	0.0147	\\
2456435.50	&	17.7734	$\pm$	0.0168	&	16.8754	$\pm$	0.0156	&	17.0539	$\pm$	0.0176	&		$\cdot\cdot\cdot$		&	$\cdot\cdot\cdot$			\\
2456435.51	&	17.7757	$\pm$	0.0197	&	16.8456	$\pm$	0.0138	&	17.0262	$\pm$	0.0147	&			$\cdot\cdot\cdot$	&	$\cdot\cdot\cdot$			\\
2456446.57	&	17.8406	$\pm$	0.0179	&	16.8872	$\pm$	0.0166	&	17.1504	$\pm$	0.0195	&		$\cdot\cdot\cdot$		&	$\cdot\cdot\cdot$			\\
2456446.58	&	17.8631	$\pm$	0.018	&	16.913	$\pm$	0.0148	&	17.1702	$\pm$	0.0205	&	$\cdot\cdot\cdot$			&		$\cdot\cdot\cdot$		\\
2456450.25	&	17.7719	$\pm$	0.0165	&	16.8362	$\pm$	0.0118	&	17.0005	$\pm$	0.0219	&	$\cdot\cdot\cdot$			&		$\cdot\cdot\cdot$		\\
2456450.25	&	17.7858	$\pm$	0.0127	&	16.8474	$\pm$	0.011	&	17.053	$\pm$	0.0153	&	$\cdot\cdot\cdot$			&		$\cdot\cdot\cdot$		\\
2456450.57	&	17.8697	$\pm$	0.016	&	16.8832	$\pm$	0.0148	&	17.1661	$\pm$	0.02		&		$\cdot\cdot\cdot$		&		$\cdot\cdot\cdot$		\\
2456450.58	&	17.8438	$\pm$	0.0144	&	16.8946	$\pm$	0.0149	&	17.1921	$\pm$	0.0219	&	$\cdot\cdot\cdot$			&	$\cdot\cdot\cdot$			\\
2456460.19	&	17.8451	$\pm$	0.0125	&	16.878	$\pm$	0.0153	&	17.1444	$\pm$	0.018	&	$\cdot\cdot\cdot$			&	$\cdot\cdot\cdot$			\\
2456460.20	&	17.8663	$\pm$	0.0124	&	16.8816	$\pm$	0.0132	&	17.1245	$\pm$	0.0286	&			$\cdot\cdot\cdot$	&	$\cdot\cdot\cdot$			\\
2456506.21	&	18.0043	$\pm$	0.0236	&	16.9962	$\pm$	0.0146	&	17.4327	$\pm$	0.0223	&	$\cdot\cdot\cdot$			&		$\cdot\cdot\cdot$		\\
2456506.21	&	18.0269	$\pm$	0.0151	&	17.0162	$\pm$	0.0099	&	17.4566	$\pm$	0.0273	&	$\cdot\cdot\cdot$			&		$\cdot\cdot\cdot$		\\
2456506.22	&	17.9992	$\pm$	0.0171	&	16.9226	$\pm$	0.0144	&	17.3753	$\pm$	0.0233	&	$\cdot\cdot\cdot$			&		$\cdot\cdot\cdot$		\\
2456506.22	&	18.0095	$\pm$	0.0158	&	16.9099	$\pm$	0.014	&	17.4377	$\pm$	0.0204	&	$\cdot\cdot\cdot$			&		$\cdot\cdot\cdot$		\\
2456796.62*	&		$\cdot\cdot\cdot$		&	18.86	$\pm$	0.2	&		$\cdot\cdot\cdot$			&		$\cdot\cdot\cdot$		&		$\cdot\cdot\cdot$		\\
2457042.35*	&		$\cdot\cdot\cdot$		&	20.16	$\pm$	0.1	&	21.18	$\pm$	0.12			&		$\cdot\cdot\cdot$		&	20.8	$\pm$	0.06	\\
2457450.50*	&		$\cdot\cdot\cdot$		&	20.61	$\pm$	0.1	&	$\cdot\cdot\cdot$				&	$\cdot\cdot\cdot$			&	$\cdot\cdot\cdot$			\\
2457824.60*	&		$\cdot\cdot\cdot$		&	20.84	$\pm$	0.1	&	$\cdot\cdot\cdot$				&	$\cdot\cdot\cdot$			&	$\cdot\cdot\cdot$			\\
\hline
\footnotesize{*R mags}
\end{tabular}\label{tab:photometry}
\end{table*}

\begin{table}
\caption{Spitzer Observations}
\begin{tabular}{@{}cccc}\hline\hline
 Date& Day    & 3.6 $\mu$m & 4.5 $\mu$m   \\ 
& (discovery) & $\mu$Jy & $\mu$Jy \\   \hline
2014-10-14&630&742$\pm$41&812$\pm$42\\
2015-05-22&850&677$\pm$40&830$\pm$43\\
\hline
\end{tabular}\label{tab:spit}
\end{table}

\subsection{Optical Spectra}
We obtained 1 epoch of low-resolution and 4 epochs of moderate-resolution optical longslit spectra with IMACS on Magellan/Baade on 2013 April 17, 2014 May 19, 2015 January 1, 2016 March 4, and 2017 March 11 (see Table \ref{tab:pcyg}).  The first observation 84 days after explosion was taken with the 300 l mm$^{-1}$ grating and 3 $\times$ 120s exposures, while the later epochs used the 1200 l mm$^{-1}$ grating centred on H$\alpha$ and 3 $\times$ 400s exposures.  Table \ref{tab:pcyg} reports the slit-width used for each observation, as well as resolving power.  Standard reductions were carried out using IRAF on a chip-by-chip basis. Flux calibration was achieved using spectrophotometric standards at a similar airmass taken with each science frame.

\begin{figure*}
\begin{minipage}[c]{0.68\textwidth}
\includegraphics[width=5.25in]{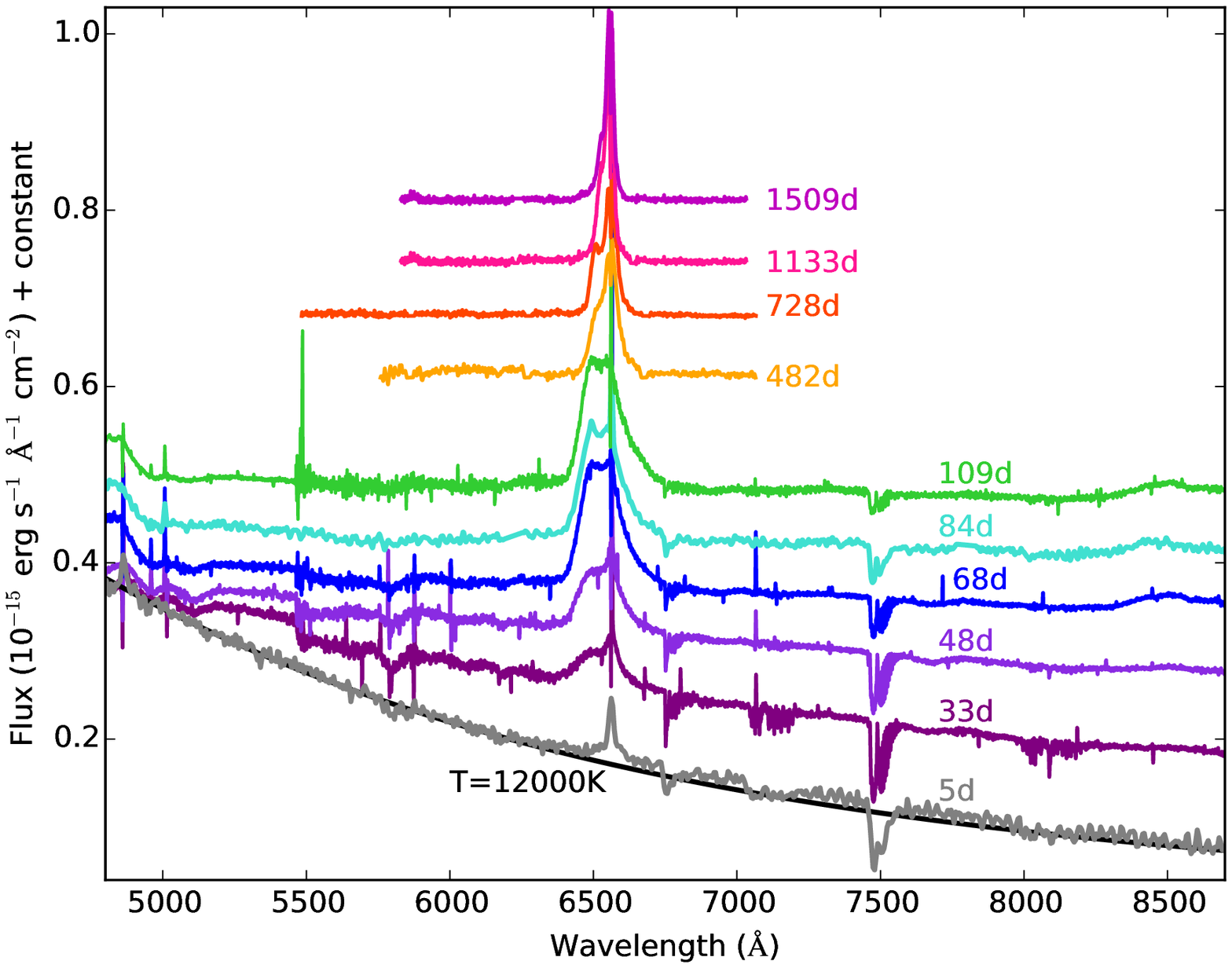}
\end{minipage}
\begin{minipage}[c]{0.31\textwidth}
\includegraphics[width=2.4in]{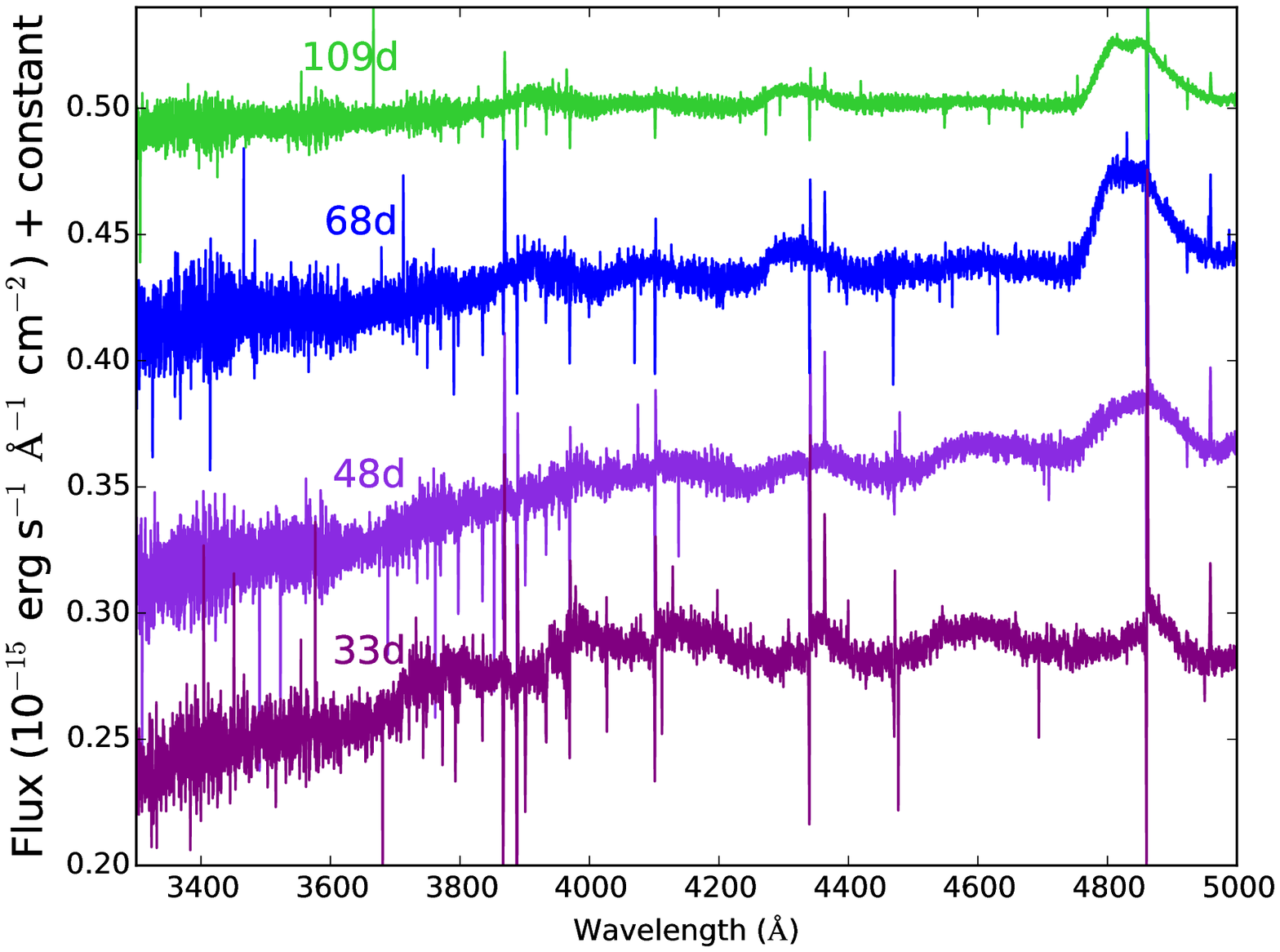}
\\[-3mm]
\includegraphics[width=2.4in]{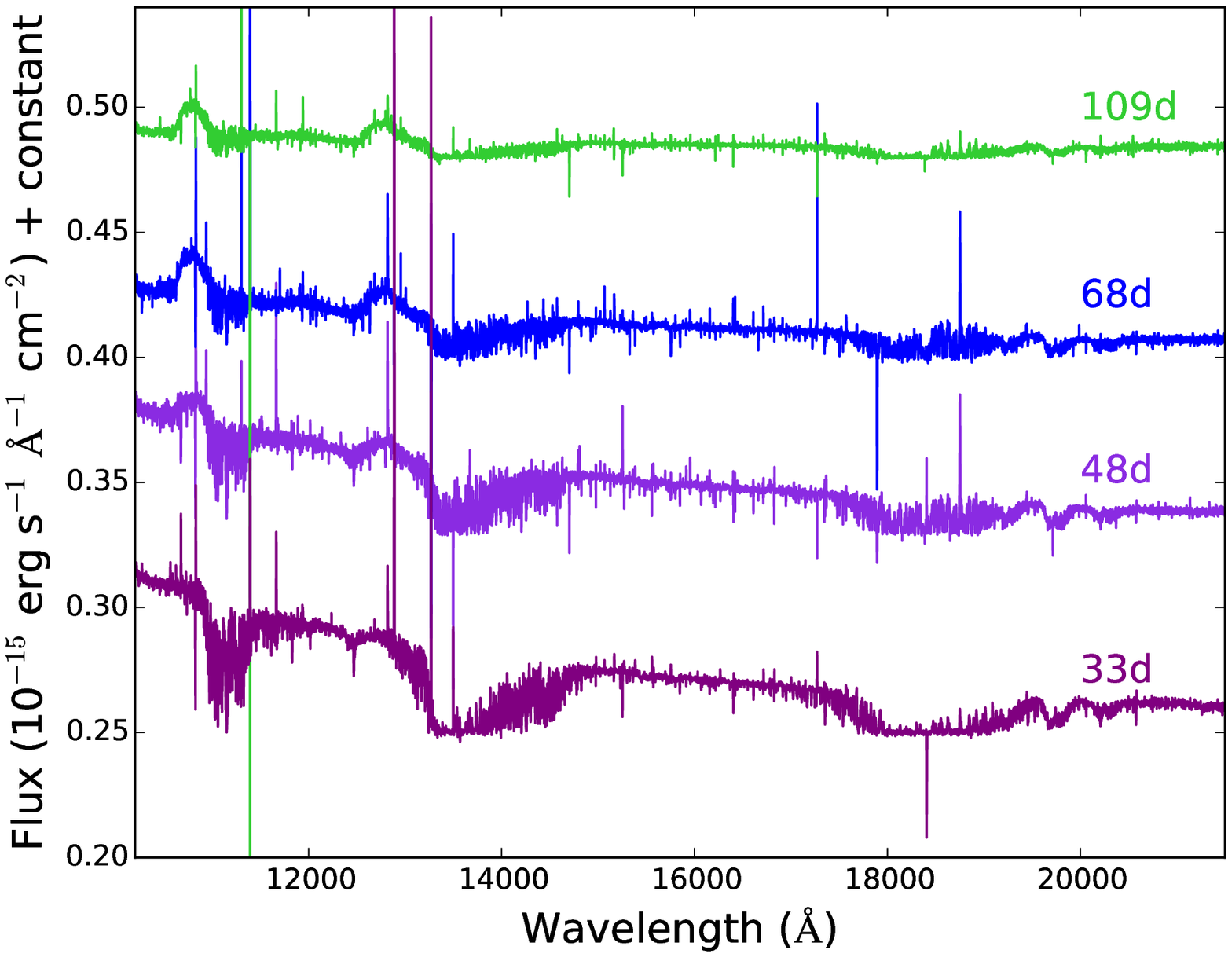}
\end{minipage}
\caption{Spectroscopic evolution of SN 2013L from 5 days to 1509 days after discovery. A summary of observational parameters is given in Table \ref{tab:pcyg}. All spectra have been corrected for Milky Way reddening of E(B-V) = 0.11, and have been shifted into the rest frame and are plotted in flux units plus a constant.  A 12000K blackbody has been plotted in black under the discovery spectrum for comparison.  The XSHOOTER NUV and NIR data are plotted on the right.  }
\label{fig:fullspec}
\end{figure*}

We have also obtained the discovery spectrum cited in \citet{2013ATel.4767....1M} taken with ESO New Technology Telescope at La Silla on 2013 January 27, using EFOSC2 and made available via the PESSTO\footnote{Public ESO Spectroscopic Survey of Transient Objects, www.pessto.org} campaign \citep{2015A&A...579A..40S}, and found on the WISeRep\footnote{http://wiserep.weizmann.ac.il/} archive \citep{2012PASP..124..668Y}. Additionally we have included VLT/XSHOOTER spectroscopy from GTO Program 090.D-0719(A), PI Hjorth made available as a Phase 3 data product on the ESO archive. Full spectral evolution of all epochs are shown in Figure \ref{fig:fullspec}.

\section{Photometric Evolution}

\subsection {Optical Lightcurve}

Figure \ref{fig:fulllc} presents the full photometric evolution of SN 2013L in g$^{\prime}$r$^{\prime}$i$^{\prime}$BV, with the first 120 days shown in detail in the inset. The lightcurves are reminiscent of Type IIL supernovae, with the first 50 days or so of photometric evolution in all filters being approximately linear, at a decline rate of 0.03 mag d$^{-1}$ in r$^{\prime}$ and V.  Around day 60, r$^{\prime}$ dramatically slows to a decline rate of  0.003  mag d$^{-1}$ which lasts for the next 140 days, until we lose the SN behind the Sun.  During this same time period, the other filters slow their decline rate as well, but are not as flat as  r$^{\prime}$, which includes the H$\alpha$ emission produced by CSM interaction.  After day 200 the sparse observations indicate a steady drop of around 3 magnitudes over a 500 day period, which then slows to roughly 0.3 mag yr$^{-1}$.   At no time do the g$^{\prime}$,r$^{\prime}$,i$^{\prime}$ and V filters follow the radioactive $^{56}$Co decay rate of $\sim$0.01 mag d$^{-1}$.  This requires additional luminosity from late-time CSM interaction.

\begin{figure}
\includegraphics[width=3.6in]{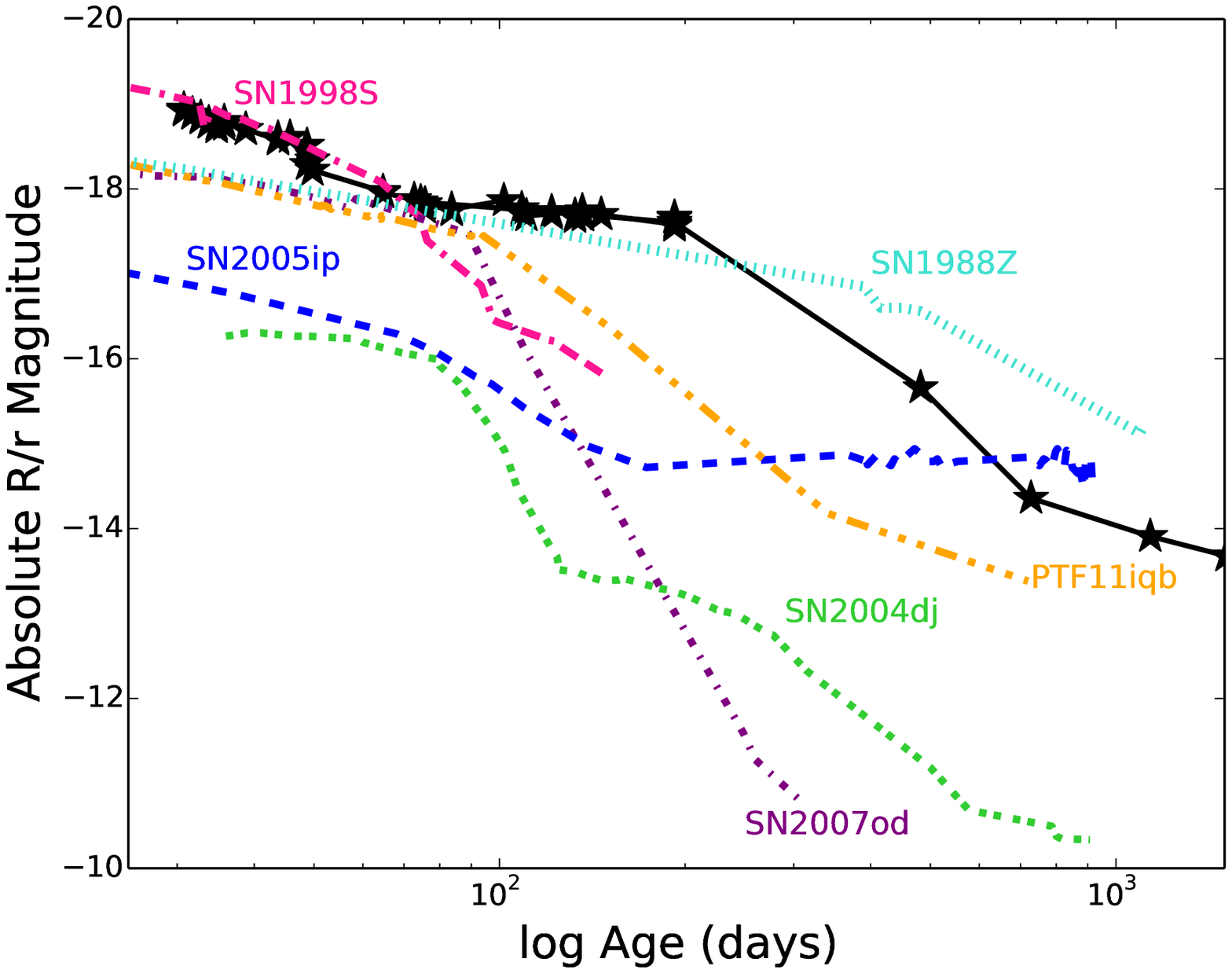}
\caption{Absolute R and r$^{\prime}$ lightcurves of CSM interacting CCSNe compared with SN 2013L (black stars).   Data are from \citet[SN2005ip]{2009ApJ...695.1334S},  \citet[PTF11iqb]{2015MNRAS.449.1876S}, \citet[SN1998S]{2000MNRAS.318.1093F}, \citet[SN2004dj]{2006AJ....131.2245Z}, \citet[SN2007od]{2010ApJ...715..541A}, and \citet[SN1988Z]{1993MNRAS.262..128T}. All data have been corrected for distance and extinction using the E(B-V) quoted in the corresponding papers, and using a standard reddening law of R$_{v}$ \citep{1989ApJ...345..245C}.}
\label{fig:LC}
\end{figure}

In Figure \ref{fig:LC} we illustrate the distinctive optical evolution of SN 2013L by comparing the absolute r$^{\prime}$/R lightcurve  with those of other interacting SNe. At early times SN 2013L most resembles SN 1998S (a IIn SNe with IIL-type lightcurve) and has a significantly brighter absolute magnitude than the IIn SN2005ip and IIP SN2004dj.   From days 60-90 post-discovery the magnitude and decline rate of SN 2013L mirrors SN 1988Z, PTF11iqb, and SN 2007od, the later of which is a "normal" SN IIP.  After day 100 the lightcurves diverge as the shock interaction begins to take over as the main energy source of SN 2013L.

Between 100--200 days the absolute magnitude of SN 2013L stays abnormally bright at around -17.7 mag, comparable only with the enduring IIn SN 1988Z which is still showing signs of interaction today \citep{2017MNRAS.466.3021S}.  Only after day 200 does the luminosity begin to fade below that of SN 1988Z, and by 2 years post-explosion the absolute R magnitude of SN 2013L resides between that of SN 2005ip and PTF11iqb (a similar pattern in H$\alpha$ is observed and discussed below.) At all times the light curve of SN 2004dj, a Type IIP SN with circumstellar interaction, is significantly fainter than the other SNe,  due to the lower density CSM surrounding a IIP explosion.

\subsection{Colour Evolution}
\begin{figure}
\includegraphics[width=3.6in]{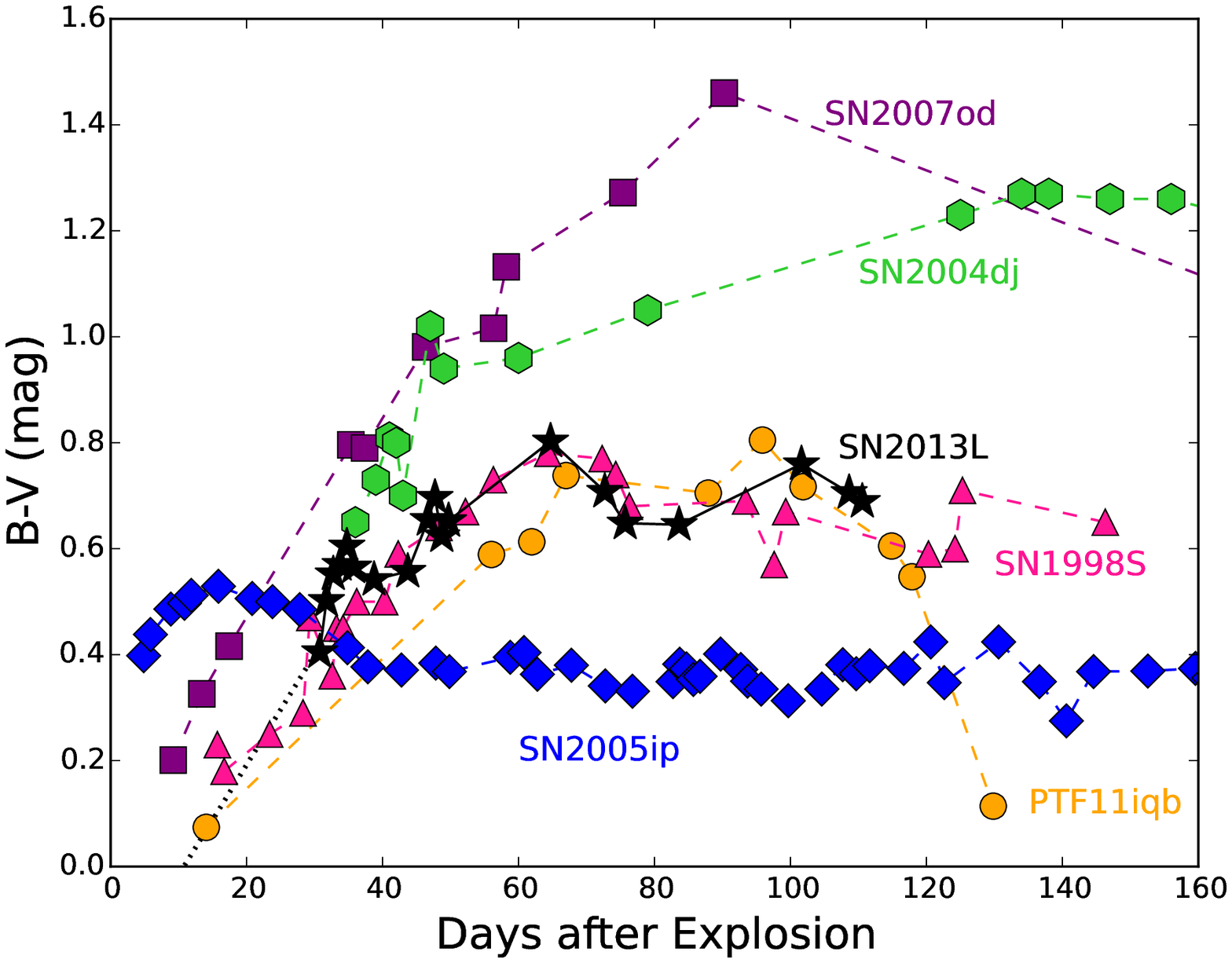}
\caption{B-V colour evolution of SN 2013L (black stars) compared with other CSM interacting IIn and IIP SNe. The dashed black line indicates a projected colour using the blackbody temperature of SN 2013L on day 5.  Data are from \citet[SN2005ip]{2012ApJ...756..173S},  \citet[PTF11iqb]{2015MNRAS.449.1876S}, \citet[SN1998S]{2000MNRAS.318.1093F},  \citet[SN2007od]{2010ApJ...715..541A}, and \citet[SN2004dj]{2006AJ....131.2245Z}. All data have been corrected for extinction using the E(B-V) quoted in the corresponding papers, and using a standard reddening law of R$_{v}$ \citep{1989ApJ...345..245C}. }
\label{fig:colorev}
\end{figure}

A look at the colour evolution of SN 2013L  further highlights the similarities with both SN 1998S and PTF11iqb (Figure \ref{fig:colorev}).   While the photometric coverage is sparse prior to day 35 an extrapolation suggests the colour started quite blue, likely B--V $\sim$ -0.1 on day 5, using the blackbody temperature of 12000K and the method presented in \citet{2012EL.....9734008B}. The colour then gradually evolved redward before leveling off at a value of B--V = 0.7, a trend identical to the aforementioned objects. What is intriguing is that the spectra of SN 2013L show persistent narrow lines, indicative of normal and long-lasting IIn SNe, while the other two only briefly show them, yet all three exhibit the same colour evolution.  Of course one tie that does seem to bind these three objects together is the presence of asymmetric emission line profiles in the intermediate width and broad components of H lines, which we will discuss in detail in the next section.

\begin{table}
\begin{center}\begin{minipage}{3.3in}
      \caption{Optical Spectroscopy of SN~2013L}
\begin{tabular}{@{}lccccc}\hline\hline
  Date    &Telescope    &Day   &Slit Width & R       \\ 
  (y-m-d)     &Instrument& &    (arcsec)  &$\lambda$/$\delta\lambda$        \\   \hline
2013-01-27 &NTT/EFOSC2  &5     & 1.0  & 300  \\
2013-02-24 &VLT/XSHOOTER  &33     & 0.9  & 7400  \\
2013-03-11 &VLT/XSHOOTER  &48     & 0.9  & 7400 \\
2013-03-31 &VLT/XSHOOTER  &68     & 0.9  & 7400 \\
2013-04-16 &Magellan/IMACS &84     & 0.7   & 1400 \\
2013-05-11 &VLT/XSHOOTER  &109     & 0.9  & 7400  \\
2014-05-19 &Magellan/IMACS &482     & 0.7    & 5300 \\
2015-01-20 &Magellan/IMACS &728     & 0.7    &  5300 \\
2016-03-05 &Magellan/IMACS &1133     & 0.7    & 5300 \\
2017-03-11 &Magellan/IMACS &1509     & 0.7    & 5300 \\
\hline
\end{tabular}\label{tab:pcyg}
\end{minipage}\end{center}
\end{table}

The other two interacting Type IIP SNe shown for comparison, SNe 2007od and 2004dj, also start out quite blue but B--V increases to $>$ 1.2 rather quickly. This pattern is the norm for most Type IIP SNe, and as the CSM interaction of these objects is delayed, is not surprising.  SN 2005ip on the other hand stays consistently around B--V = 0.4 mag, a pattern also seen in SN 1988Z and SN 2010jl \citep{1993MNRAS.262..128T,2012AJ....144..131Z}.  Considering the late-time endurance of the light curve of SN 2013L, one could expect the colour evolution to mimic these objects with very strong CSM interaction and slow fading, yet it is redder and likely evolved that way rather quickly.

A quick examination of Figure \ref{fig:colorev} shows that we could roughly subdivide these Type II events into three classes, the red Type IIP, the blue and flat IIn, and the intermediate 98S-like.  The normal IIP supernovae quickly go from blue to red as the the large RSG envelope expands and cools. The classic IIn like SN2005ip on the other hand stay rather blue as the shock interaction keeps the temperature high, and often a blue pseudocontinuum, caused by a blend of narrow fluorescing metal lines in the surrounding CSM, can appear \citep{2009ApJ...695.1334S}.  This would suggest that the intermediate IIn objects show a mix of the SN photosphere and CSM interaction, which likely is caused by asymmetry in the CSM. The spectral evolution of these objects (SN 1998S, PTF11iqb, and SN 2013L) in fact all show spectral signatures of an asymmetric CSM as will be discussed more below.

\subsection{Infrared Evolution}
Spitzer IRAC 3.6 and 4.5 $\mu$m imaging from days 630 and 850 (Figure \ref{fig:spitzimage}) show a bright source at the location of SN 2013L. As we show in Table 2 the SN emission is  roughly 0.8 mJy, relatively bright for a SN so distant and of that age.  For comparison the 3.6 $\mu$m flux for SN 2010jl on day 621 was 8.63 mJy \citep{2014ApJ...797..118F} and for SN 2005ip on day 948 was 5.76 mJy \citep{2011ApJ...741....7F}.   If moved to the distance of SN 2013L (72 Mpc) these would correspond to 4.0 and 1.4 mJy respectively. Both IIn SNe have components of pre-existing and newly formed dust
 \citep{2010ApJ...725.1768F,2011AJ....142...45A,2012AJ....143...17S,2014Natur.511..326G,2009ApJ...691..650F,2009ApJ...695.1334S}, which could also be the case for SN 2013L.  Whether this emission is from an IR echo, shock heated pre-existing dust, or from newly formed dust grains can be difficult to disentangle.  Unfortunately no earlier Spitzer observations exist, so it is impossible to compare how the IR emission has evolved with time.  We also cannot discount a nearby or underlying source as the cause of some or all of the emission, as no pre-SN imaging exists.  This is unlikely because of the high luminosity ($>$ 2.5 $\times$ 10$^{7}$ L$_{\sun}$) and a change in flux between the two observations presented here.  Continued monitoring with Spitzer (and eventually JWST) will allow us to decipher more precisely the IR environment surrounding SN 2013L.

\begin{figure}
\includegraphics[width=3.6in]{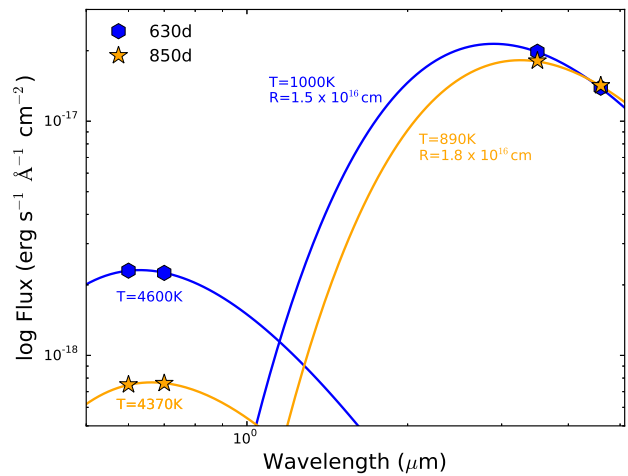}
\caption{SED of SN2013L on day 630 and 850.  The IR points are from the Spitzer observations while the optical data is the continuum emission on either side of H$\alpha$ extrapolated from the day 482, 728, and 1133 optical spectra.  The solid lines are the blackbody fits to both the optical and IR components of each epoch, and the corresponding temperatures and blackbody radius are indicated on the plot.}
\label{fig:SED}
\end{figure}

A blackbody fit to the two Spitzer passbands can give us a rough idea of the temperature and location of the dust.  Figure $\ref{fig:SED}$ shows the best fits to the Spitzer epochs on days 630 and 850.  Both dates can be fit with temperatures of 1000 K and 890 K respectively, a reasonable temperature for all three scenarios.  The blackbody radius (R$_{bb}$) can give us a stringent lower limit of the location of the IR luminosity.  R$_{bb}$ grows from 1.5 to 1.8 $\times$ 10$^{16}$ cm between the two dates, or from 5 to 7 light days.  By day 630 the initial flash of the SN has traveled well beyond 1.7 light years, but this does not necessarily discount the presence of a light echo scattering off of intervening gas and dust.  It does however take a velocity of 2750 km s$^{-1}$ and 2450 km s$^{-1}$ to travel the distance of R$_{bb}$ in 630 and 850 days respectively.  These velocities are almost identical to the central velocity of the blue-component seen in our H$\alpha$ line around the same time (Figure $\ref{fig:halpha}$ and Table $\ref{tab:fwhm}$.)  This strongly advocates for pre-existing CSM that is being shock-heated by the expanding SN ejecta.

 The dates, and more importantly, the temperatures of the IR observations are conducive to the formation of dust grains so we also cannot completely rule out new dust formation in the ejecta or post-shock gas.  As we will discuss further in Section 4, we do believe the case for only newly formed dust is a bit weak, as the emission line asymmetries are most likely caused by the SN geometry rather than just dust attenuating the receding emission.

\section {Spectroscopic Evolution}
\begin{figure}
\includegraphics[width=3.6in]{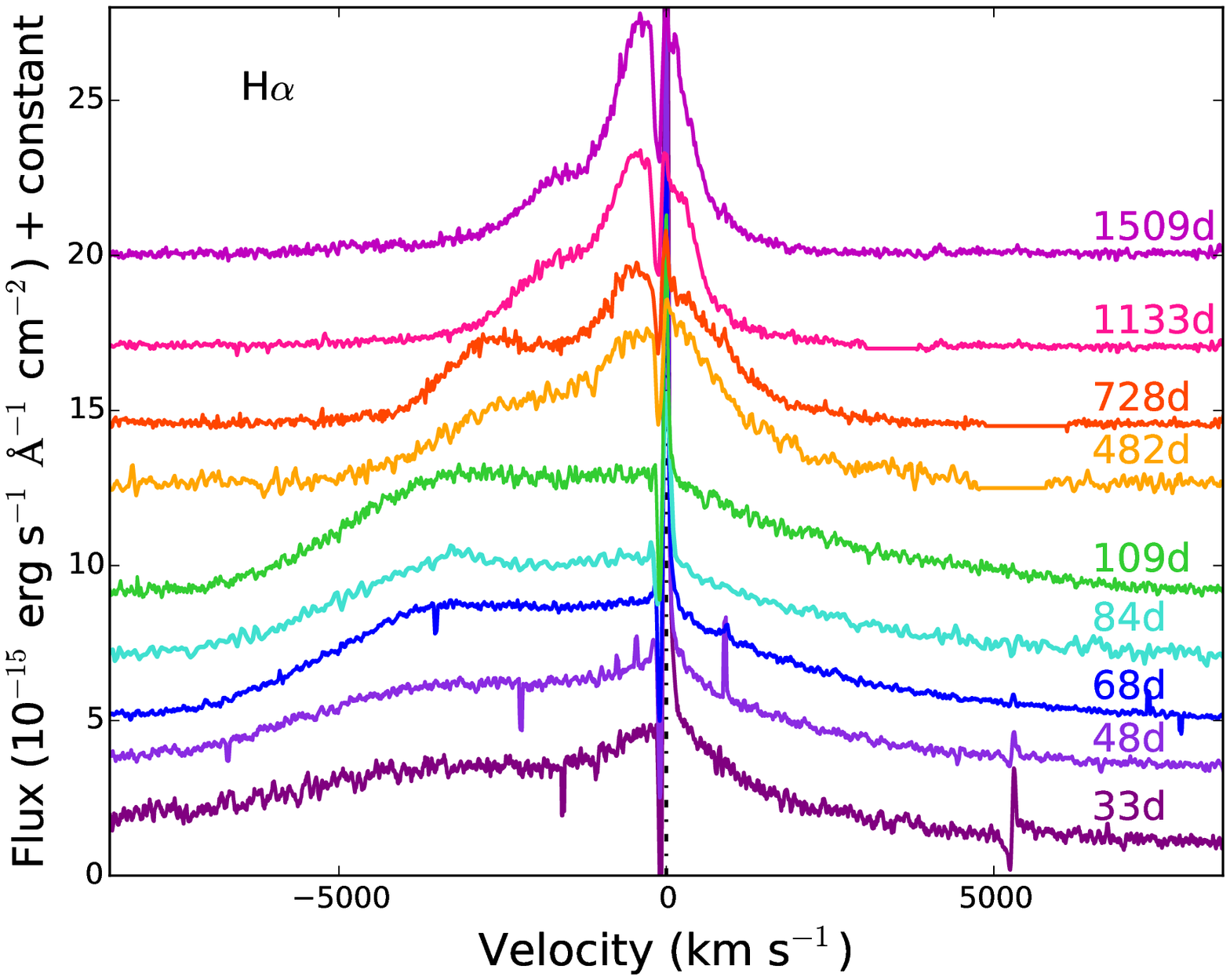}
\includegraphics[width=3.6in]{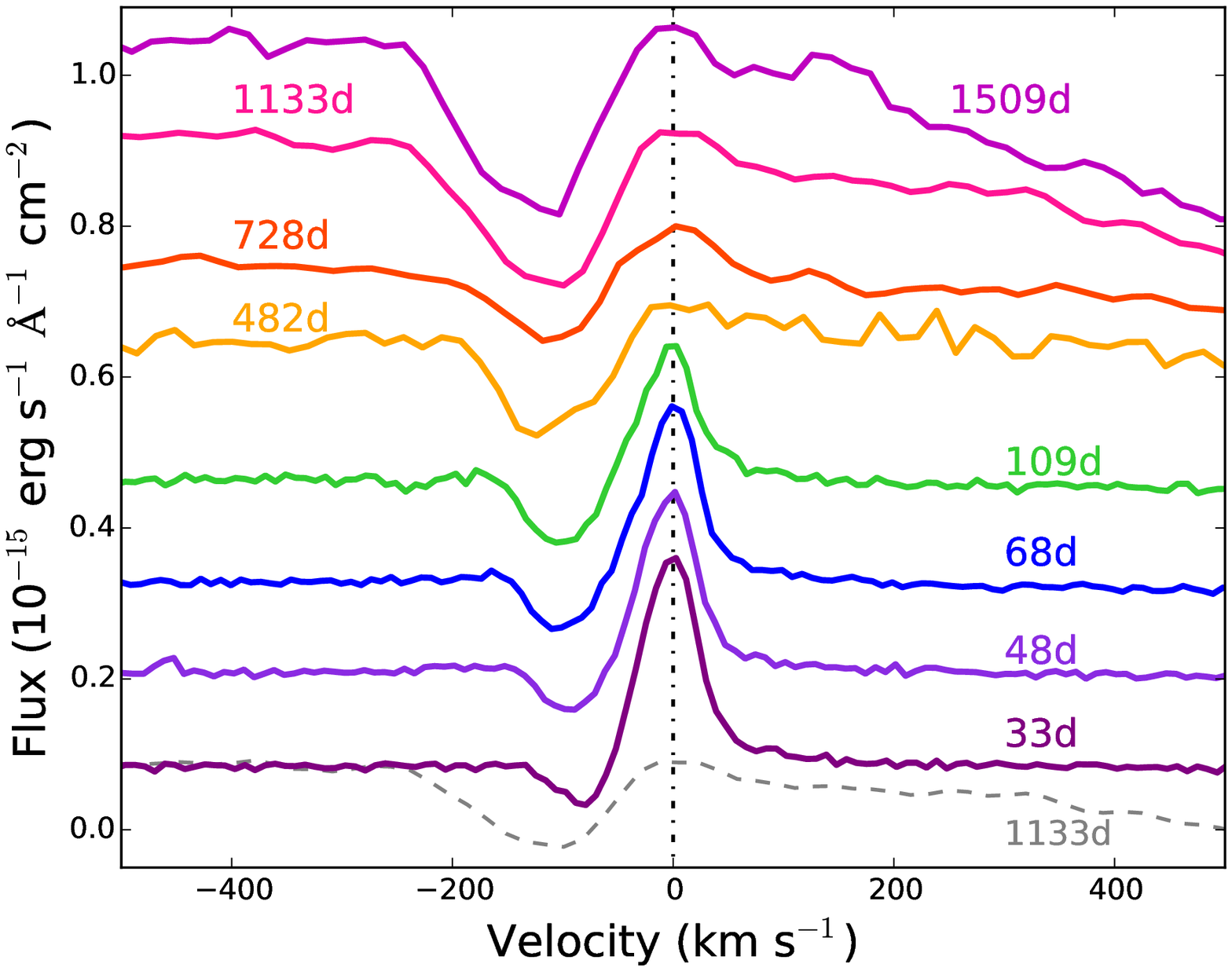}
\caption{Spectral evolution of H$\alpha$ in velocity space.  The \textbf{top} panel shows the intermediate width emission lines and the evolution of the blue-bump with time, while the \textbf{bottom} panel is zoomed to show the evolution of the narrow  emission and absorption components. The comparison between the narrow line shape on day 33 and day 1133 can be seen in the bottom panel as the  dashed grey line is the emission from day 1133 scaled and shifted to match the blue side of the line. Zero velocity is indicated by the black dash-dotted line, and has been chosen to be centred on the narrow H$\alpha$ emission. }
\label{fig:halpha}
\end{figure}

\begin{figure*}
\includegraphics[width=3.4in]{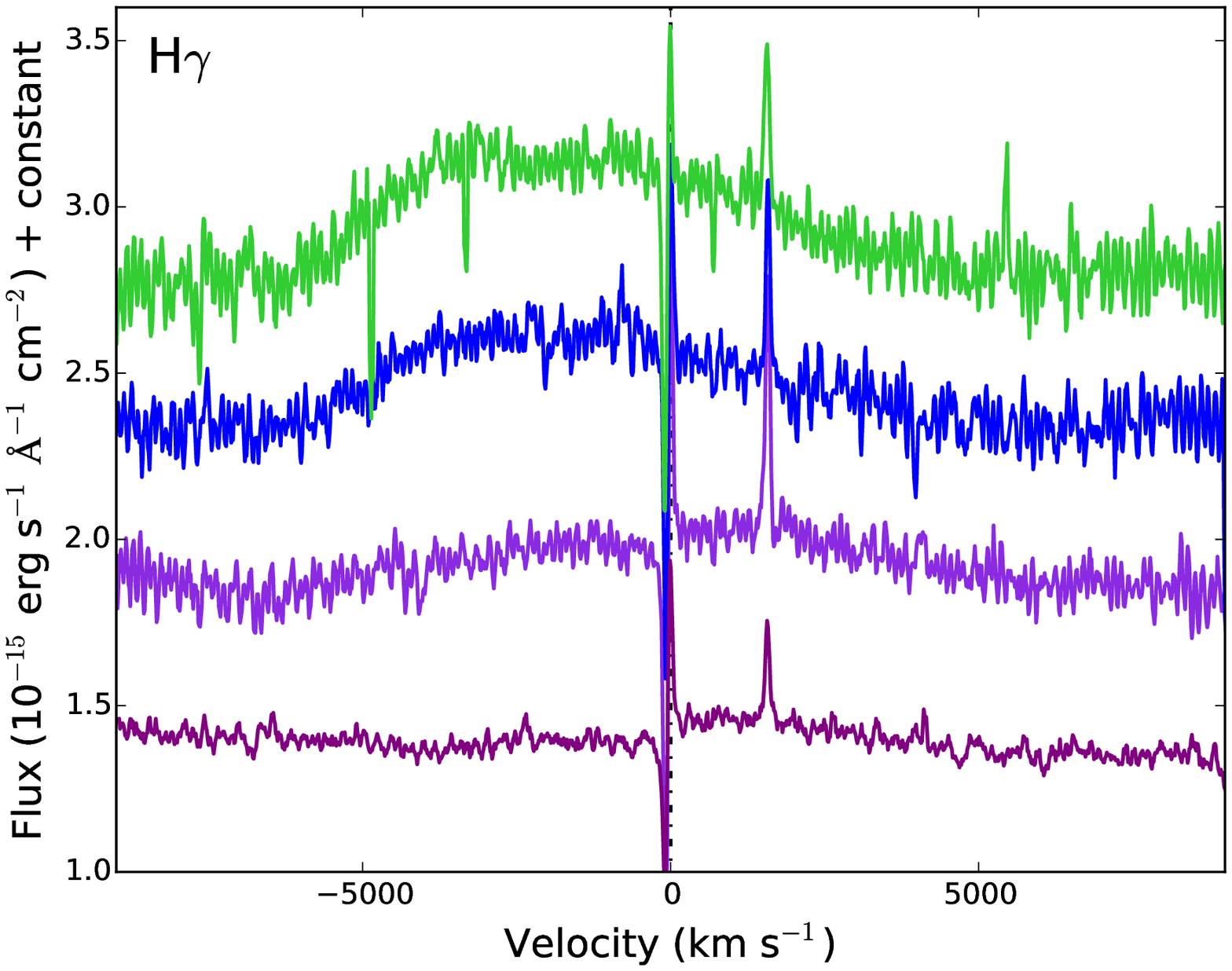}
\includegraphics[width=3.4in]{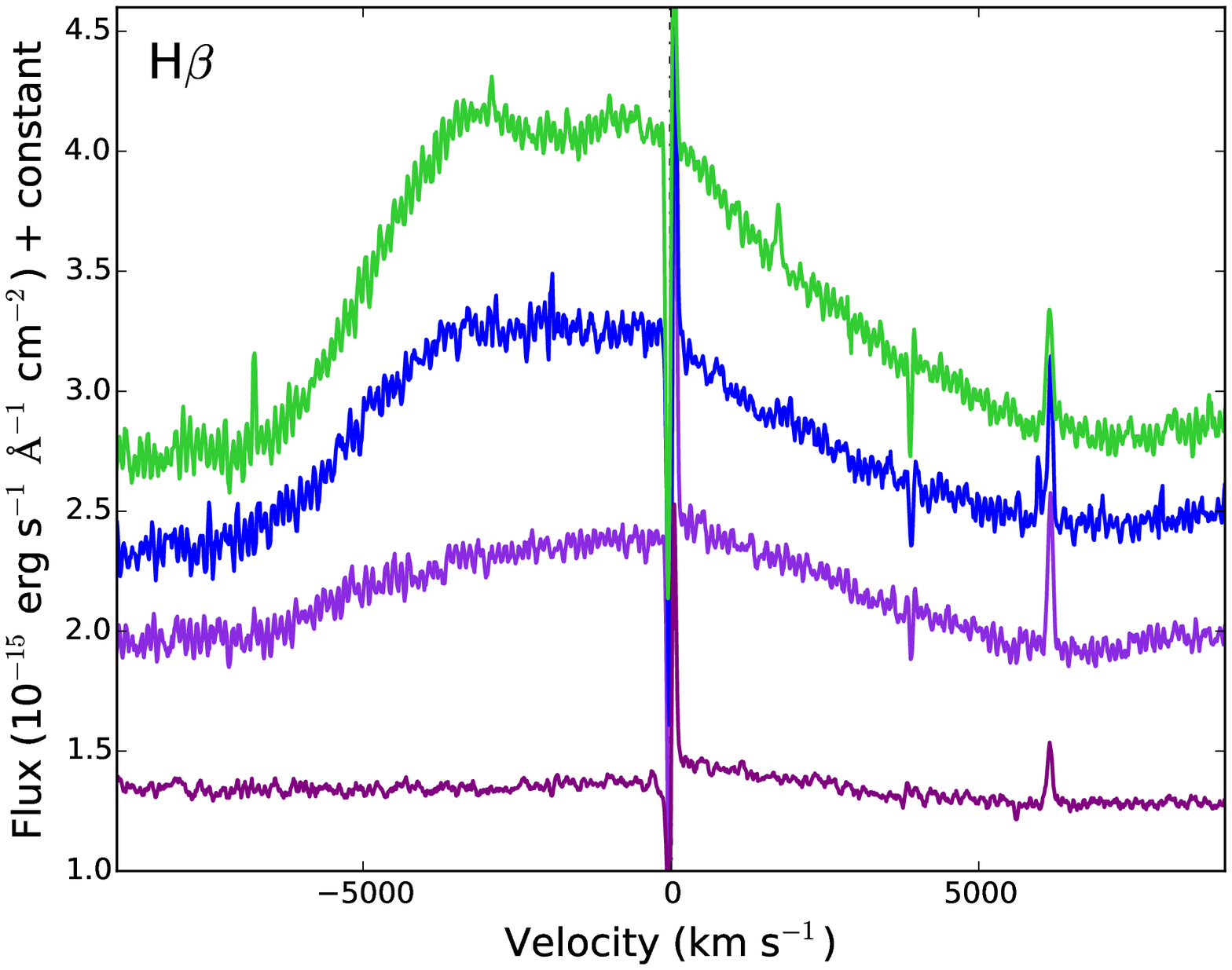}
\includegraphics[width=3.4in]{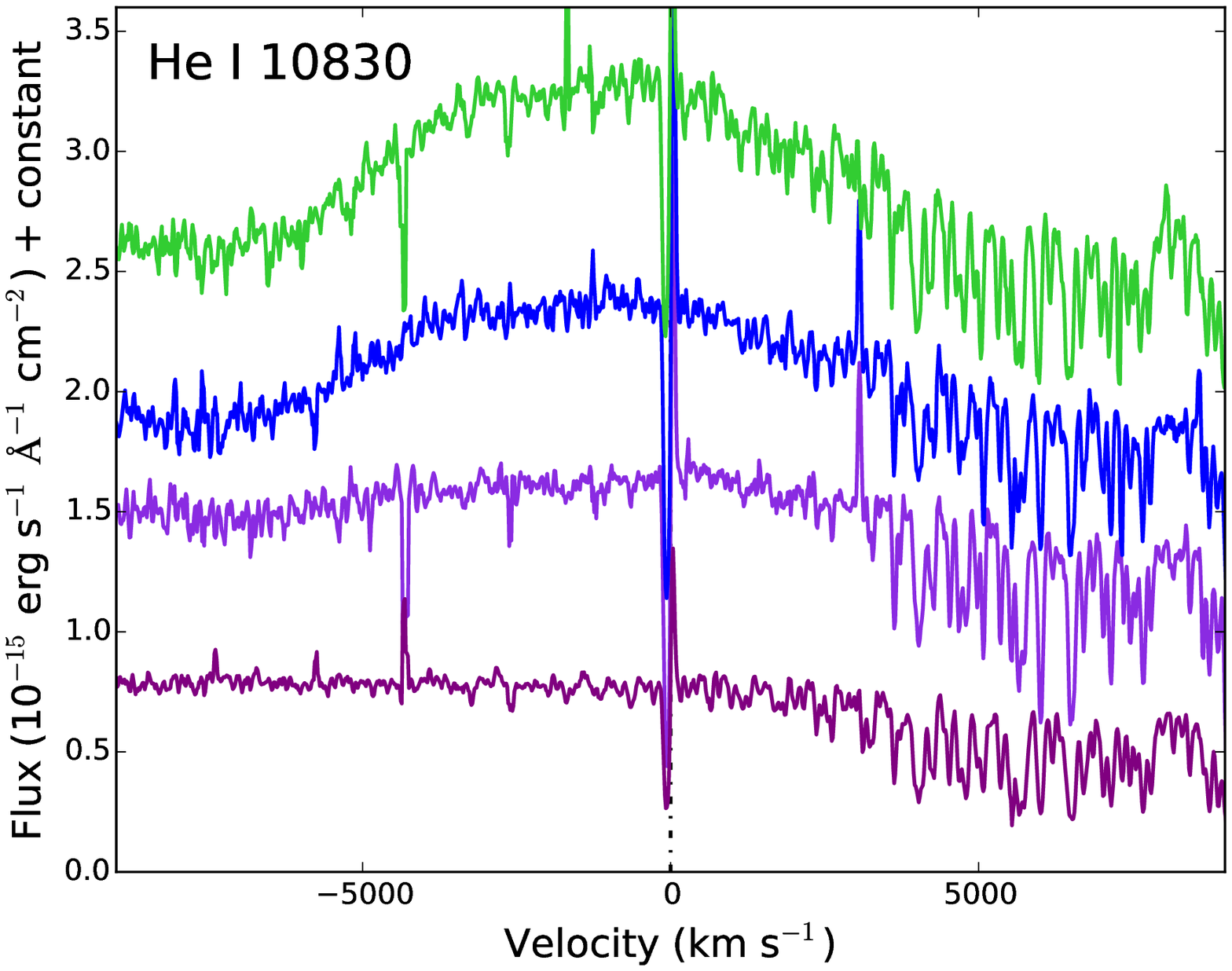}
\includegraphics[width=3.4in]{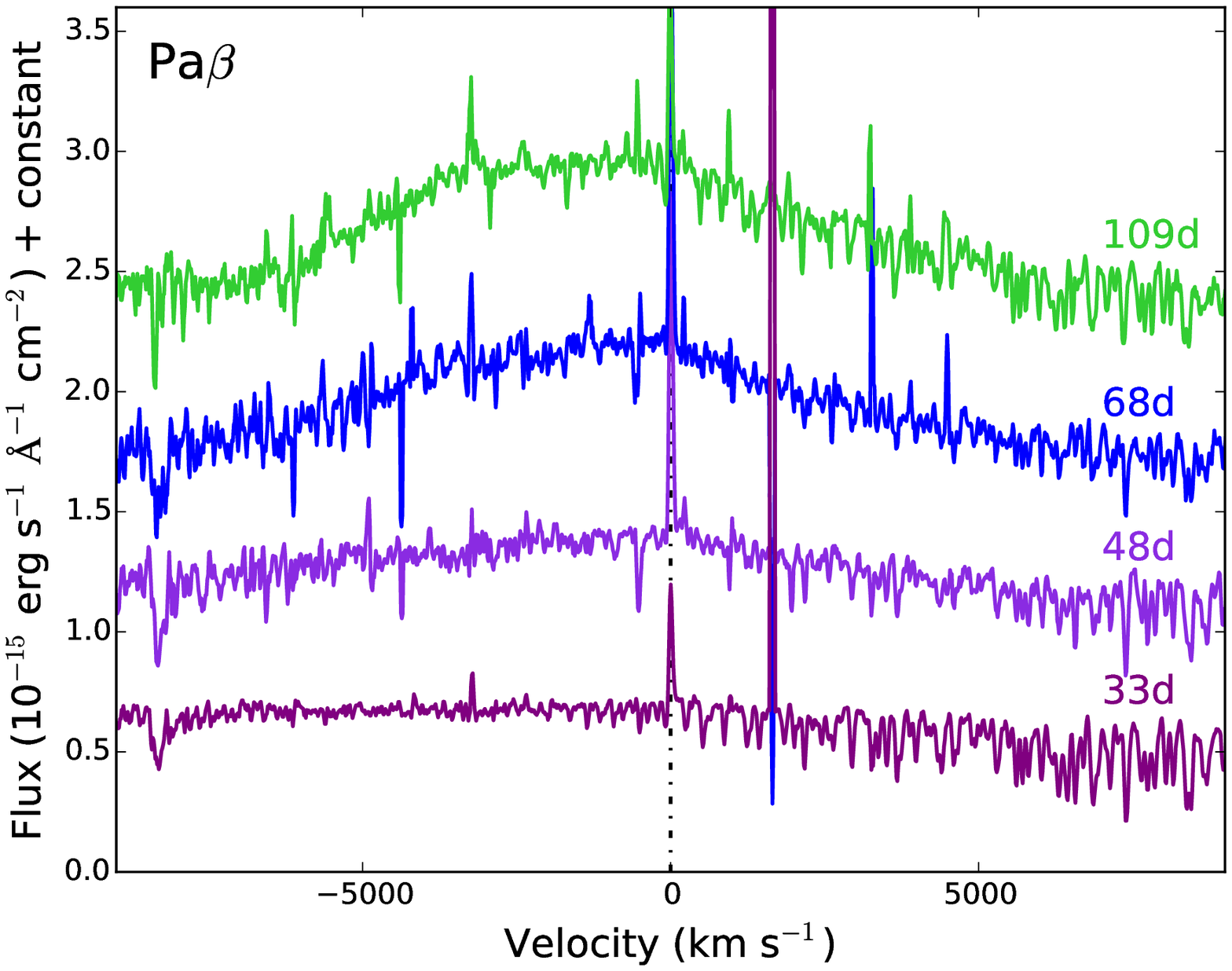}
\caption{Same as Figure \ref{fig:halpha}, but in velocity space zoomed in on various hydrogen or helium emission lines which are indicated in the top corner of the plot.  Only the epochs with the highest resolution are included in this plot, and all NIR emission is from only the XSHOOTER spectra. Zero velocity is indicated by a dashed vertical line.}
\label{fig:narrow}
\end{figure*}

The first spectrum of SN2013L on day 5 after discovery, shown in grey at the bottom of Figure \ref{fig:fullspec}, shows a mostly featureless continuum with only narrow hydrogen Balmer lines present in emission.  A blackbody fit yields a temperature of  12000 K and a photospheric radius of 7 $\times$ 10$^{14}$ cm, similar to other Type IIn SNe at around the same age  \citep{2015MNRAS.449.1876S,2016MNRAS.456..323K}. There were no WR lines seen in the earliest spectrum, like was observed for other IIn SNe 1998S \citep{2015ApJ...806..213S}, SN 2013cu \citep{2014Natur.509..471G}, and PTF11iqb \citep{2015MNRAS.449.1876S}. This does not of course discount their presence at earlier epochs.

Narrow P-Cygni absorption and emission of the Balmer lines and He I 10830 \AA\ can be seen in the spectrum on day 33. Narrow emission was also present on day 5, but the low resolution of that spectrum makes the detection of an absorption feature impossible. In all but the H$\alpha$ line, the absorption component is stronger than the emission.  Narrow emission is also seen in the Paschen and Brackett series in the NIR, with no evidence of an absorption component.  Similarly, the He I lines in the optical only show narrow emission.  

\begin{figure}
\includegraphics[width=3.6in]{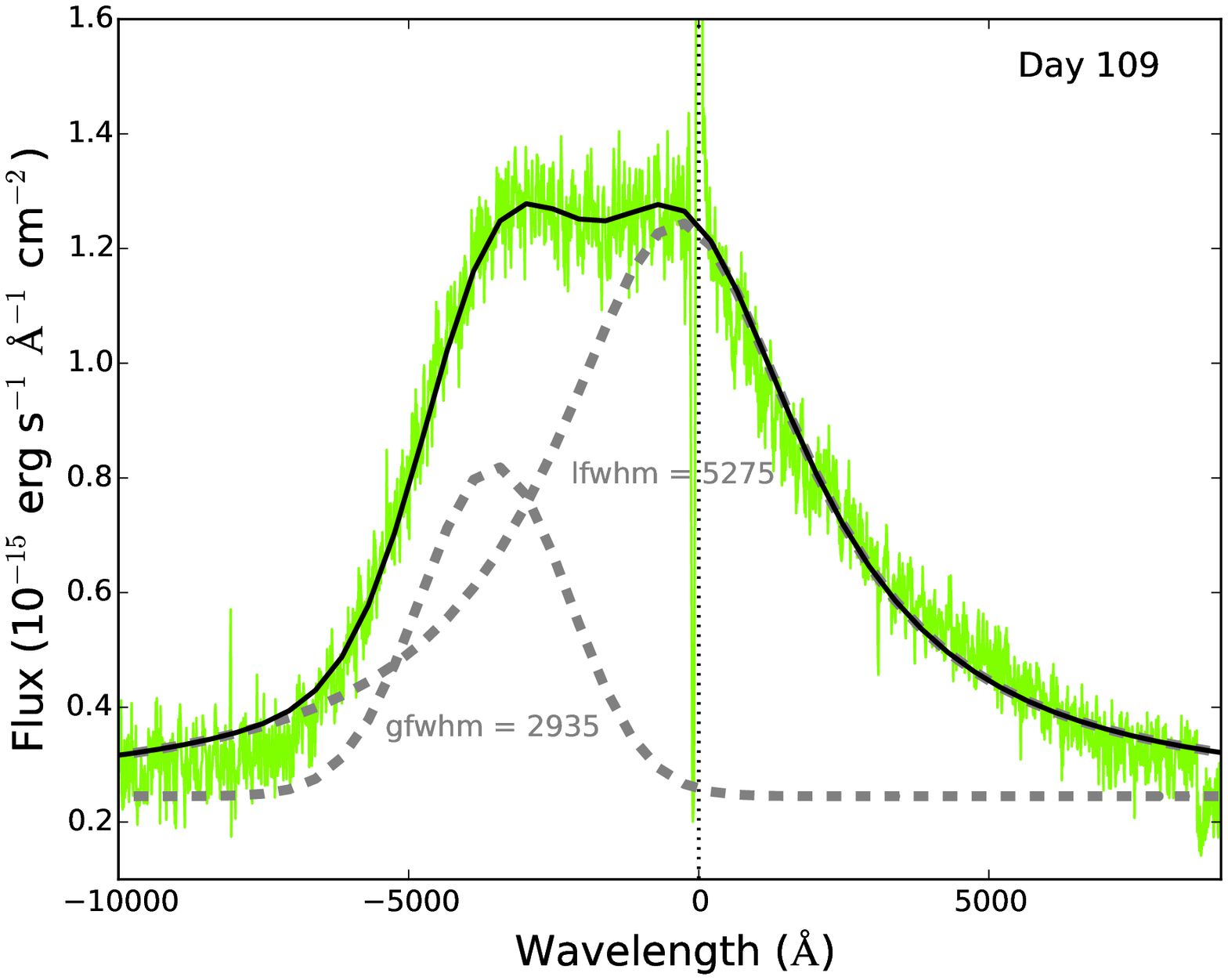}
\includegraphics[width=3.6in]{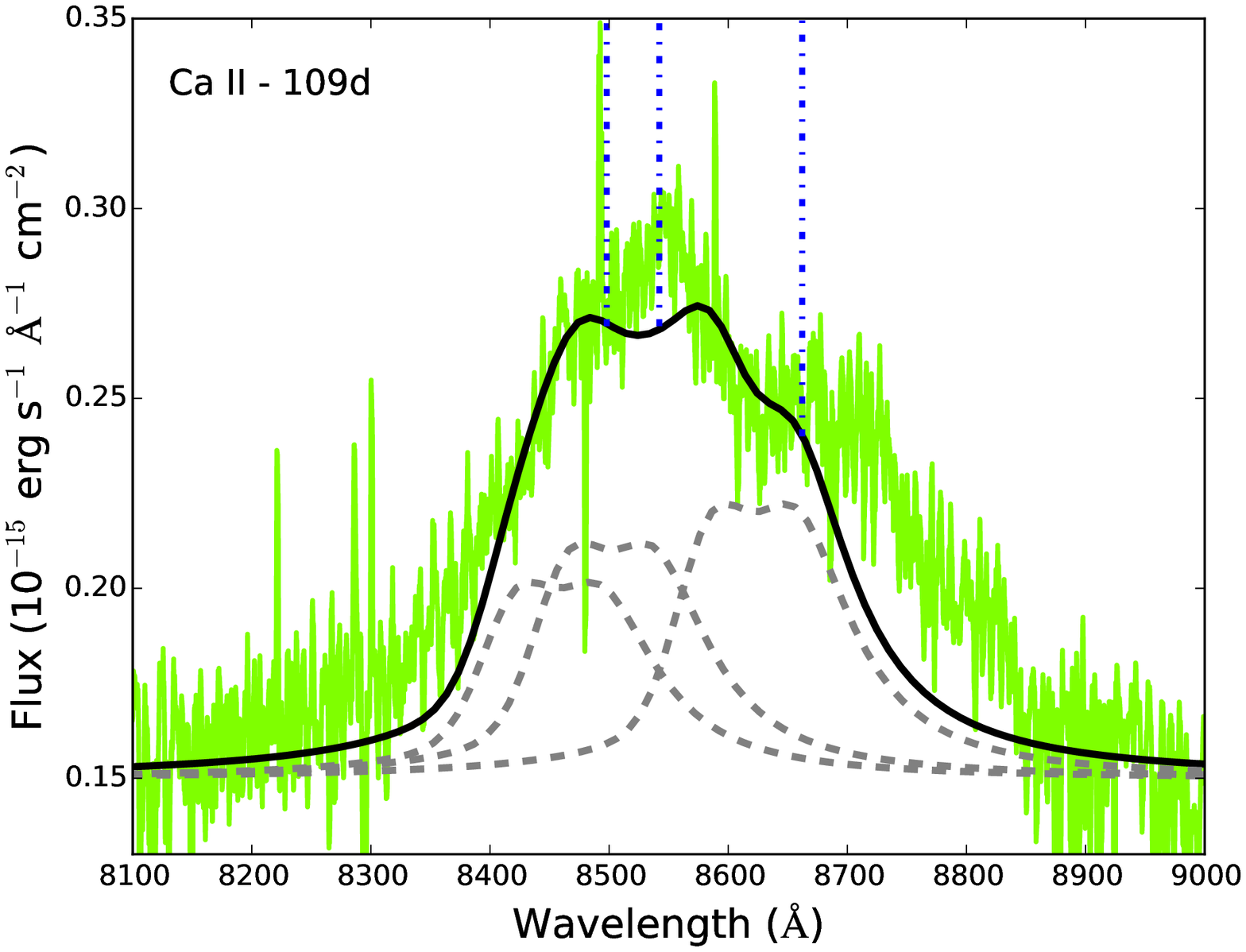}
\caption{\textbf{Top:} Multi-component fit  (black line) to the  broad H$\alpha$ emission line on day 109 (green).  Allowing the narrow emission to reside at zero velocity the profile can be recreated with a Lorentzian centred at  -337 km s$^{-1}$ and a Gaussian blue shifted to $\sim$ 3515 km s$^{-1}$, both of which are shown by the grey dashed line. \textbf{Bottom:} The Ca II $\lambda\lambda\lambda$ 8498,8542,8662 \AA\ IR triplet from day 109 (green) with a fit created from scaling and combining the fit from the H$\alpha$ profile seen above for each of the three lines.  Each individual profile is shown in the grey dashed line. }
\label{fig:components}
\end{figure}

\begin{figure}
\includegraphics[width=3.6in]{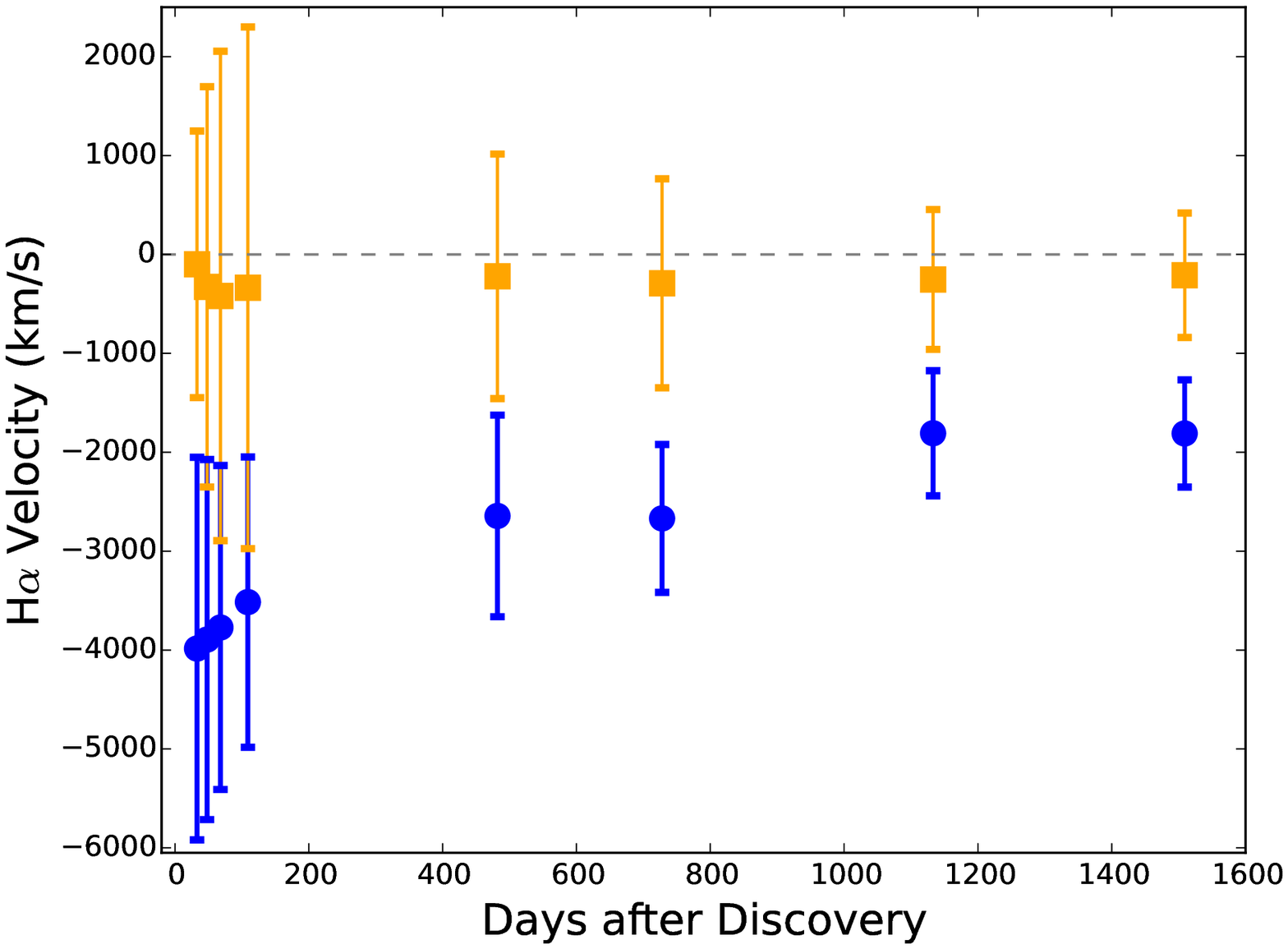}
\caption{Evolution of the centroid and FWHM of the Lorentzian (orange) and Gaussian (blue) emission of H$\alpha$ with time. The centroid is shown as the square or circle symbol, and the FWHM by the velocities spanned by the vertical bars.}
\label{fig:fwhm}
\end{figure}

The Balmer and He I 10830 \AA\ lines (with the addition of Pa$\beta$) also show intermediate-width components with FWHMs of 2000 - 4000 km s$^{-1}$, with one centred at roughly zero velocity  and the other  centred a few thousand km s$^{-1}$ to the blue (Figure \ref{fig:halpha} and Figure \ref{fig:narrow}). A comparable red emission feature is not seen.  We can fit the H$\alpha$ profiles with three distinct features: a narrow emission profile with prominent P-Cygni absorption, an intermediate width Lorentzian centred near zero velocity, and an intermediate width Gaussian profile centred at roughly -3500 km s$^{-1}$. The broad Lorentzian profile  occurs when optically thick material causes multiple electron-scattering events.  The resulting narrow part of the profile traces the expansion velocity of the emitting material, in this case the slow moving progenitor wind or the CSM, while the broad wings are a function of the optical depth of the material \citep{2001MNRAS.326.1448C}.    Figure \ref{fig:components} shows the two intermediate-width components needed to create the multipeaked profile of H$\alpha$ on day 109.  Both the Lorentzian and the Gaussian components are blushifted from the central narrow component.  The FWHMs and other information of the components are listed in table \ref{tab:fwhm} and shown in Figure \ref{fig:fwhm}.

With time, the FWHM of the blue shifted component decreases, from roughly 3900  km s$^{-1}$ on day 33 to $\sim$ 1100 km s$^{-1}$ on our last observation on day 1509. During this same time period the centre of the blue component shifts from -4000 km s$^{-1}$ to -1800 km s$^{-1}$. The behavior of the central Lorentzian emission is a bit less straightforward, with the FWHM rising to a maximum of $\sim$ 5300 km s$^{-1}$ on day 109, and then gradually falling again to $\sim$ 1300 km s$^{-1}$ by day 1509.  The narrow emission also gets broader and fainter and by day 482 the narrow absorption component is more pronounced and is in fact stronger than the narrow emission by our last observations. The fact that we are still seeing narrow P-Cygni  absorption even on our last observation on day 1509 may provide important clues about the density and geometry of the CSM as well as our viewing angle.  Similar profiles have been seen in SN 1998S  and PTF11iqb and will be discussed in detail below.

At no time do the [O I] lines at $\lambda\lambda$6300,6363 \AA\  appear, and there is a general lack of a nebular phase typically seen in normal Type IIP SNe. This is common in SNe IIn where CSM interaction tends to mask the underlying nebular phase spectrum (see \citet{2014ARA&A..52..487S}). There does appear to be an intermediate-width feature around 8500 \AA\ which was also seen in the spectrum of SN 2010jl presented in \citet{2014ApJ...797..118F}. There it was attributed to O I $\lambda$8446 \AA\ emission, but as we show in Figure \ref{fig:components}, we can approximate the shape of the emission in SN 2013L by using the H$\alpha$ profile of the same time centred at each of the Ca II IR triplet locations ($\lambda\lambda\lambda$8498, 8542, 8662 \AA\ ). While not an exact fit, the combination of self-absorption and broad line-widths can account for the discrepancy.  No narrow lines are seen at any of the Ca II IR locations, suggesting that Ca II emission lines arise in the fast ejecta or at the location of the reverse shock.

In our last epoch on day 1509 the narrow emission seems to have all but vanished, and only the narrow absorption feature remains.   Assuming that slow RSG winds reach as far as 3 $\times$ 10$^{18}$ cm, or 3.2 ly \citep{1995MNRAS.273L..19L}, we would expect the narrow emission lines ionized by the SN blast wave to begin fading by $\sim$ 3 years. The persistence of the narrow absorption may be an indication of geometry; if there is an equatorial enhancement of the CSM we may be viewing that nearly edge-on.  Continued monitoring will reveal how that absorption feature evolves, and could potentially show the appearance  of the [O I] emission lines as the oxygen layers begin to interact with the reverse shock, as was seen in SN 1998S \citep{2012MNRAS.424.2659M}.

\begin{table}
\begin{center}\begin{minipage}{3.3in}
      \caption{FWHM and centres of H$\alpha$ Components in km s$^{-1}$}
\begin{tabular}{@{}lccccc}\hline\hline
  Epoch    & Narrow Abs & \multicolumn{2}{c}{Lorentzian} & \multicolumn{2}{c}{Gaussian }     \\ 
(days) & $\upsilon$$_{centre}$&FWHM&$\upsilon$$_{centre}$&FWHM&$\upsilon$$_{centre}$  \\   \hline

33 & -84 &2696 &-100& 3869&-3985\\
48& -91 &4047 &-328&3640&-3894\\
68&-100&4949 &-420&3275&-3772\\
109&-98&5275&-337&2935&-3515\\																																																																																																																																																																																																																																																																						482&-104&2472&-221&2039&-2644\\
728&-114&2113&-292&1495&-2669\\
1133&-126&1484&-253&1265&-1808\\
1509&-130&1258&-210&1084&-1810\\

\hline
\end{tabular}\label{tab:fwhm}
\end{minipage}\end{center}
\end{table}

\subsection{The multi-peaked shape of H$\alpha$}
One of the most striking features of SN 2013L is the multi-component structure of H$\alpha$ (and other Balmer and He I lines), in particular the additional blue shifted emission around -3500 km s$^{-1}$.  Often, a blue shifted asymmetry can be explained by dust formation attenuating the receding red side of the SN more so than the blue.  First detected in SN 1987A \citep{1989LNP...350..164L}, this type of evidence for dust formation has been seen in many CCSNe including SN 2003gd \citep{2006Sci...313..196S}, SN 2004et \citep{2009ApJ...704..306K}, SN 2005ip \citep{2010ApJ...725.1768F,2011MNRAS.412.1522S}, SN 2007od \citep{2010ApJ...715..541A}, and one of the clearest cases SN 2006jc \citep{2008ApJ...680..568S}. In conjunction with the emission line asymmetry, a decrease in the optical light curve and a corresponding increase in the IR brightness is often observed as new dust grains form in the ejecta.  No early IR data exist for SN 2013L as we mention above, and any optical obscuration in the light curve due to dust can be hidden by the ongoing CSM interaction. The lack of any emission on the red side of H$\alpha$ in all epochs, plus the presence of the blue emission by day 33 when ejecta temperatures would probably have been too high for dust condensation may suggest another mechanism for the multi-peaked lines.

\begin{figure}
\includegraphics[width=3.4in]{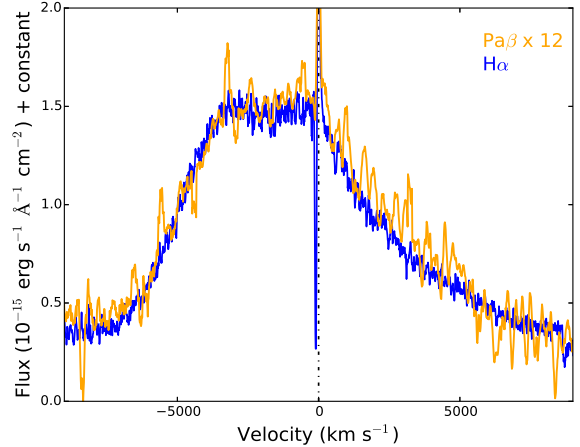}
\caption{Comparison of H$\alpha$ (blue) and Pa$\beta$ (orange) emission on day 109.}
\label{fig:dust}
\end{figure}

In Figure  \ref{fig:dust} we plot H$\alpha$ and Pa$\beta$ from our day 109 spectrum.  Adjusting the flux of the Pa$\beta$ emission line to match that of H$\alpha$ shows that there is little difference in the profile shape.  This further supports the hypothesis that extinction from newly formed dust is not the cause for the asymmetry. If it were due to dust, the red-side of H$\alpha$ would be more attenuated than that of Pa$\beta$ due to the wavelength dependence of the dust extinction.  Since the bluer H$\alpha$ shows the same shape as the redder Pa$\beta$, the asymmetry is intrinsic to the SN or the CSM.

While we cannot rule out that there is some newly-formed dust  responsible for a portion of the hydrogen asymmetry, it is more likely an asymmetry in the surrounding CSM. If we invoke a disc-like geometry of the CSM and then view it edge-on the CSM can easily obscure the red side of the emission. This is further supported by the persistent narrow P-Cygni lines, particularly the absorption feature at late times.  On day 1509 the blue emission bump is centred at -1810 km s$^{-1}$, this would mean the minimum radius for the location of the material responsible for the blue emission bump is $\sim$ 2 $\times$ 10$^{16}$ cm ($\sim$ 1330 AU) away.  Of course instead of a disc seen edge-on and therefore obscuring the receding side, we could potentially have a situation with only CSM on the blue side of the SN, similar to PTF11iqb which showed a density enhancement on the red side \citep{2015MNRAS.449.1876S}.  The one-sided geometry (i.e. the blue emission bump at -3500 km s$^{-1}$) could  be explained by a SN exploding in a binary system where mass transfer is occurring preferentially to one side due to orbital eccentricities.  Case in point is RY Scuti which is in a state of mass transfer from a binary companion and shows a resolved torrodial CSM with a density enhancement on the far side \citep{2001ApJ...559..395G,2002ApJ...578..464S,2011MNRAS.418.1959S}. A one-sided eruption is another possibility to create an asymmetric event.  For instance \citet{2016MNRAS.463..845K} show that $\eta$ Carinae experience asymmetric, one-sided eruptions in 1250 and 1550 AD, centuries prior to the Great Eruption in the 1800s.

 Asymmetric CSM is not uncommon, and over the past decade as the number of well-observed CCSNe has grown, many SNe IIn have displayed evidence of disc, torus, or bi-polar CSM.  This includes SN 1998S \citep{2000ApJ...536..239L}, SN 1997eg \citep{2008ApJ...688.1186H}, SN 2009ip \citep{2014AJ....147...23L,2014MNRAS.438.1191S,2014MNRAS.442.1166M}, SN 2010jl \citep{2011A&A...527L...6P,2011AJ....142...45A,2012AJ....143...17S,2015ApJ...810...32C}, PTF11iqb \citep{2015MNRAS.449.1876S}, and SN 2015bh \citep{2016MNRAS.463.3894E}.  Recently \citet{2017ApJ...834..118M} suggested that the IIP/L SN2013ej SN likely had an equatorial enhanced CSM close to edge on based on spectropolarimetry. Similarly SN 2004dj \citep{ 2006AstL...32..739C}, SN 2007od \citep{2010ApJ...715..541A}, SN2011ja \citep{2016MNRAS.457.3241A} and other type IIP/L also show observational signatures of asymmetric CSM (Figure \ref{fig:speccomp}).  While not exhibiting the narrow line emission, the late-time H$\alpha$ of these IIP/L SNe was multi-peaked, suggesting that a continuum of CSM interaction strengths and geometries exists.  SN 2013L may be similar to those, but it also happens to have closer and denser CSM which made it appear as a Type IIn.

\section{DISCUSSION}

\subsection{The enduring luminosity of H$\alpha$}
\begin{figure*}
\includegraphics[width=5.5in]{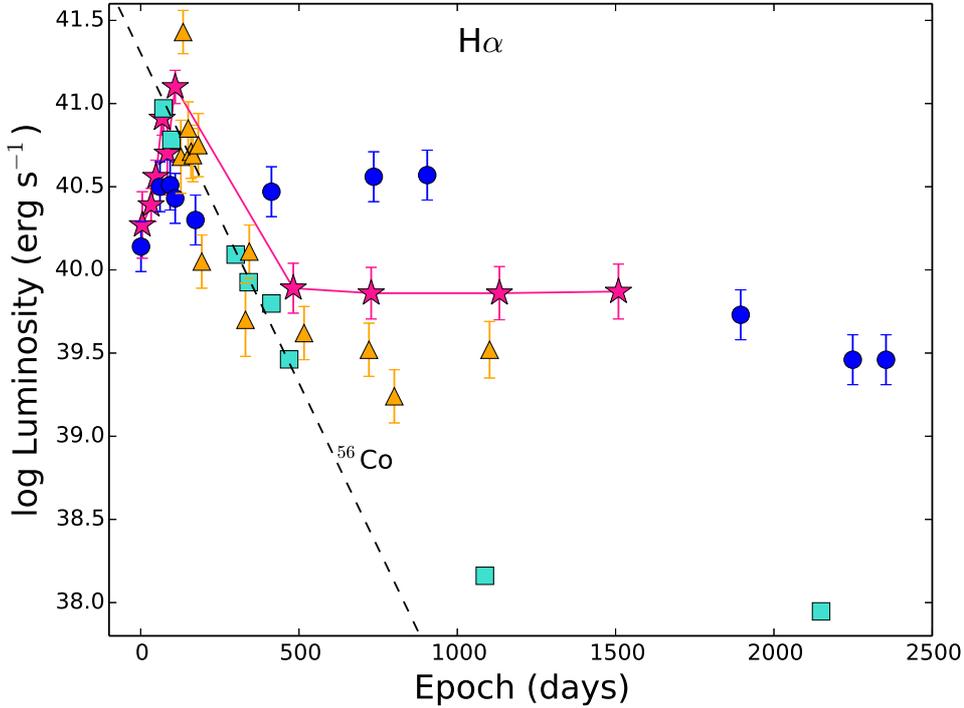}
\caption{H$\alpha$ luminosity of SN2013L (pink stars) compared with other  IIn SNe. Data for 1998S (teal squares) comes from Figure 4 of \citet{2012MNRAS.424.2659M}, for PTF11iqb (orange triangles) from Figure 9 of \citet{2015MNRAS.449.1876S}, and for SN2005ip (blue circles) from \citet{2017MNRAS.466.3021S}. The radioactive decay luminosity of $^{56}$Co is plotted as a dashed black line.   }
\label{fig:halum}
\end{figure*}

The late-time optical emission of SN 2013L is powered by the SN ejecta colliding with the CSM, which is clearly illustrated in the most recent spectroscopic observations.  From day 482 and onwards, the optical spectra are dominated by strong  H$\alpha$ emission.  To illustrate the strength of this late-time emission, in Figure  \ref{fig:halum} we have plotted the total integrated H$\alpha$ luminosity of SN 2013L from each of the epochs listed in Table 3, along with other well-studied and enduring IIn SNe.  This figure is adapted from \citet{2017MNRAS.466.3021S}, and \citet{2012MNRAS.424.2659M}.  SN 2013L starts with L$_{H\alpha}$ similar to that of SN 2005ip, but quickly rises to the levels of PTF11iqb.  The emergence of the blue-bump in the Balmer lines by day 33 and the plateau in the R-band light curve after day $\sim$ 50 suggests that  the ejecta began interacting with even denser CSM from the very start of explosion. There are no spectra available between days 109 and 482, so it is possible the luminosity continued to rise before declining down to the level detected over a year later, or as is suggested by Figure  \ref{fig:halum}, it could have declined at roughly the radioactive  $^{56}$Co decay rate.

By day 482 the L$_{H\alpha}$  is much brighter than would be expected from radioactivity alone, and remains at roughly this brightness even on our last observation on day 1509. This is evidence of  interaction powered H$\alpha$.  On day 1133, SN 2013L is  $\sim$45 $\times$ brighter than SN 1998S at the same age, and $\sim$2.5 $\times$ brighter than PTF11iqb.  There is a lack of data for SN 2005ip during this time-period, but we can estimate SN 2013L was likely $\sim$2-3 $\times$ fainter.  This would suggest that SN2013L has a wind density roughly halfway between SN 2005ip and PTF11iqb, or a mass-loss rate between 1 $\times$ 10$^{-4}$ and 1 $\times$ 10$^{-3}$ M$_{\sun}$ yr$^{-1}$ \citep{2015MNRAS.449.1876S,2009ApJ...695.1334S,2017MNRAS.466.3021S}.  Further comparisons among these three objects will be explored below.

Although we do not have X-ray or Radio data of SN 2013L, we can use the luminosity and kinematics of the system offered to us through our optical spectra to determine a wind-density parameter and therefore a mass-loss rate of the progenitor star.  This is expressed as $\frac{\dot{M}}{V_{CSM}}$ = $\frac{2L}{V_{SN}^{3}}$, where V$_{CSM}$ is the CSM velocity measured from the minimum of the narrow P-Cygni lines, V$_{SN}$ is the supernova expansion velocity,  and L = L$_{H\alpha}$/$\epsilon$  (see \citet{2016arXiv161202006S}).   As a lower bound we can expect $\epsilon$ $\sim$ 50$\%$, since the thermal energy produced should be about half of the kinetic energy. For the upper bound we assume a conservative L$_{H\alpha}$ = 0.1L. SN 2010jl, PTF11iqb, and SN 2009ip may have similar CSM geometries and efficiency factors used for these objects are  5$\%$--7$\%$ \citep{2014ApJ...797..118F}, 15$\%$ \citep{2015MNRAS.449.1876S}, and $<$10$\%$ \citep{2014MNRAS.442.1166M}, respectively.   On day 33,  V$_{CSM}$ = 84 km s$^{-1}$, V$_{SN}$ = 2696 km s$^{-1}$, and L$_{H\alpha}$= 10$^{40.4}$ resulting in an $\dot{M}$ = 0.34 -- 1.75 $\times$ 10$^{-3}$  M$_{\sun}$ yr$^{-1}$. This same mass-loss rate is recovered within uncertainties throughout the evolution of the SN, and by day 1509 has increased to $\sim$  1.5--7.6 $\times$ 10$^{-3}$ M$_{\sun}$ yr$^{-1}$, consistent with the mass-loss rates of PTF11iqb and SN 2005ip as hypothesized above. Combined with wind velocities between 80--130 km s$^{-1}$ this points to either a YHG or a non-eruptive LBV progenitor for SN 2013L \citep{2016arXiv161202006S,2012ApJ...744...10K}.

\subsection{Progenitor Characteristics}

As reported in Table \ref{tab:pcyg}, the resolving power of the XSHOOTER observations is $\sim$ 7400, or 40 km s$^{-1}$.  The measured P Cygni absorption of the narrow emission component is $\sim$ 100 km s$^{-1}$, larger than the resolution of the data.  Therefore we can assume that we are resolving the line and that the CSM surrounding the SN has a wind speed of 80--130 km s$^{-1}$ depending on the epoch.  This wind velocity places the progenitor in a YHG or LBV wind regime which suggests a M$_{ZAMS}$ $>$ 25 M$_{\sun}$.

As has been eluded to throughout the text, the H$\alpha$ profile seems to suggest disc-like geometry. Rings and discs indicating asymmetric mass-loss (perhaps due to interaction with a binary companion) are common around many evolved massive stars \citep{1997ApJ...475L..45B,2007AJ....134..846S,2017A&A...597A..99W}. The high instances of massive stars in binaries, of which 70--75$\%$ will likely interact \citep{2012Sci...337..444S, 2012ApJ...751....4K, 2014ApJ...782....7D}, make a binary system that causes equatorial mass-loss likely.  Close-binary progenitors are often invoked for Type IIb SNe \citep{1993Natur.364..509P,2011MNRAS.412.1522S,2013ApJ...762...74B,2014ApJ...790...17F,2011ApJ...739L..37M}, wherein the very nearby companion strips off the hydrogen envelope and the SN explodes as a YSG with only trace amounts of H in the atmosphere. If instead the binaries are further apart the radius of the RSG can increase as nuclear burning progresses, causing mass-loss and binary interaction \citep{2014ApJ...785...82S}.  This mass-loss would most likely be disc-like and reside in the equatorial plane. 

The high mass-loss rate, the asymmetric H lines, and the persistent narrow absorption lines are all suggestive of a disc or torus of CSM residing close the the progenitor of SN 2013L which was created due to a binary companion.  The fact that we see a IIn and not a Ib/c or IIb would seem to indicate that system was in RLOF when the progenitor exploded since far too much hydrogen is still present for substantial atmospheric stripping to have occurred. The progenitor scenario for SN 2013L could even be similar to the current YHG  star HR 5171A. This highly evolved star appears to be in Wind Roche-Lobe Overflow with a much smaller companion, wherein the YHG wind does not completely fill the RL, but it is still gravitationally stripped onto the companion \citep{2014A&A...563A..71C}.

\begin{figure}
\includegraphics[width=3.4in]{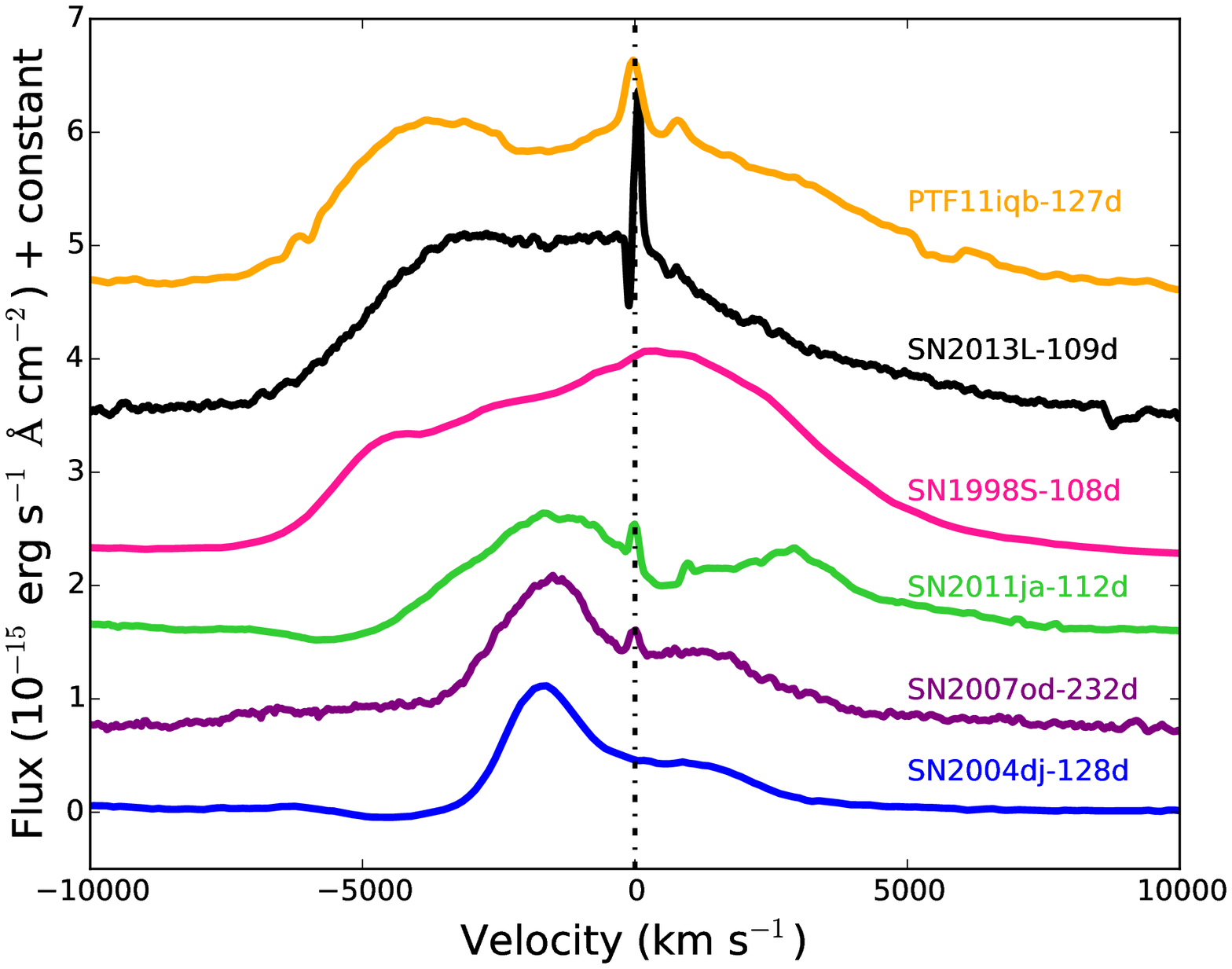}
\includegraphics[width=3.4in]{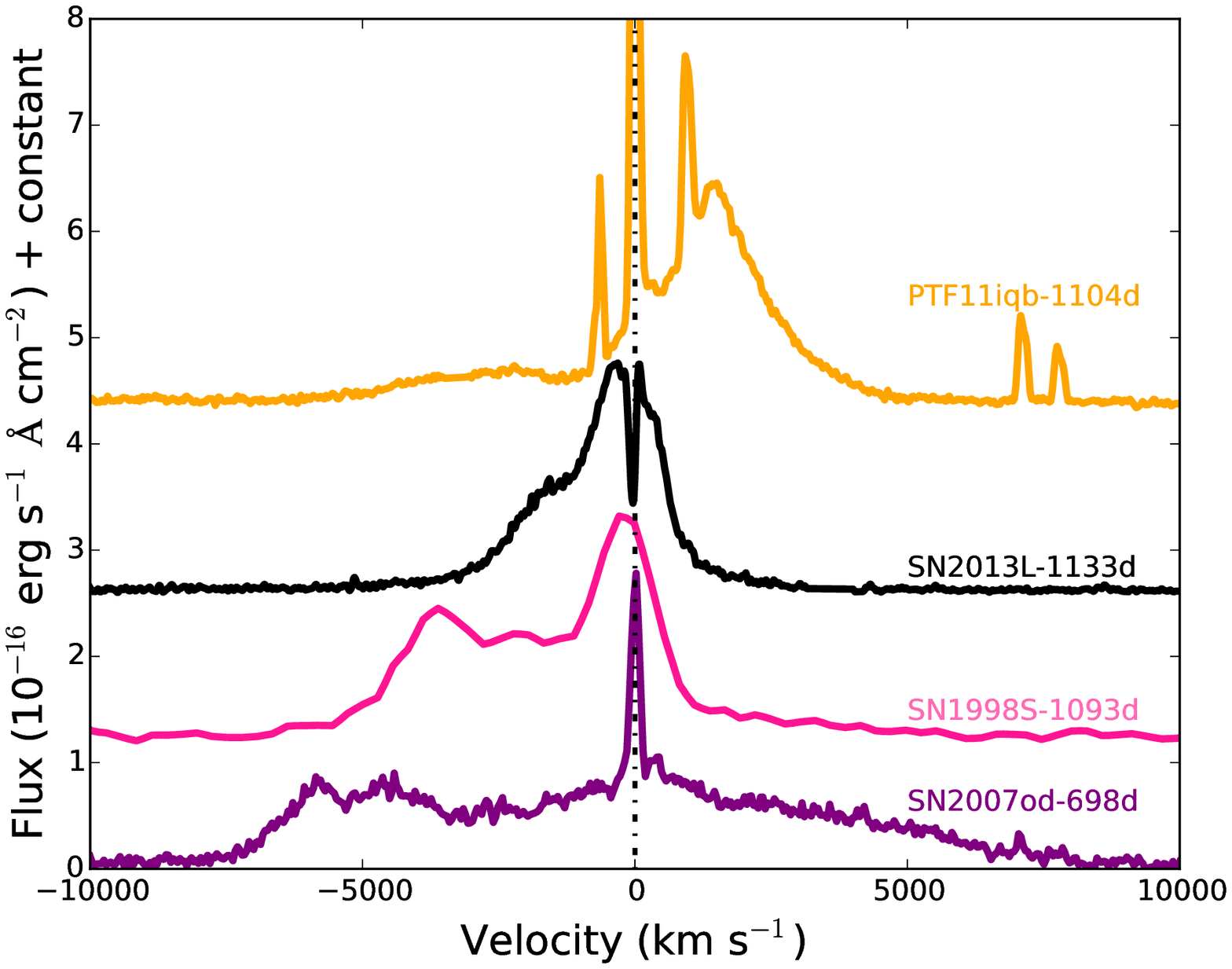}
\caption{Comparison of the H$\alpha$ profile of SN 2013L with other interacting CCSN at early (top) and late (bottom) times.  $\textbf{Top:}$ SN 2013L on day 109 in relation to three IIP SNe  2004dj  \citep{2006MNRAS.369.1780V}, 2007od \citep{2010ApJ...715..541A}, and 2011ja \citep{2016MNRAS.457.3241A} as well as two IIn SNe 1998S and PTF11iqb \citep{2015MNRAS.449.1876S} at similar ages.  $\textbf{Bottom:}$ Same as top but for day 1133 for SN 2013L as well as much later times for SNe 2007od, 1998S, and PTF11iqb. }
\label{fig:speccomp}
\end{figure}
  
\subsection{Comparisons with other IIn}
\subsubsection {SNe 1998S and PTF11iqb}

The unique profiles and persistent luminosity of H$\alpha$ in SN 2013L warrant a comparison with other CCSNe that show signs of strong CSM interaction.  As is clearly seen in Figure \ref{fig:LC}, with an
absolute magnitude of  -19.0$\pm{0.2}$, SN 2013L is on par with PTF11iqb (-18.4) and  SN 1998S (-19.5).  Not only are the absolute magnitudes similar, but the B-V colour evolution (Figure \ref{fig:colorev}) of SN 2013L, SN 1998S, and PTF11iqb are also almost identical in the first few months. Around 100 days the lightcurves  begin to diverge, with PTF11iqb and SN 1998S continuing to drop in magnitude (the former at a less drastic rate than the later) and SN 2013L hitting a plateau in R-band brightness.   From these two diagnostics alone it would seem that these three objects belong to their own subset of IIn CCSNe, but the addition of the H$\alpha$ profiles further support this hypothesis. 

Illustrated in Figure \ref{fig:speccomp}, the multiple-peaked H$\alpha$ seen in SN 2013L is reminiscent of that seen in SN 1998S and PTF11iqb. While the complex H$\alpha$ was seen starting by day 33 in SN 2013L and day 37 in PTF11iqb, it was not until $\sim$ 200 days when the multiple peaks appeared in SN 1998S (although asymmetries were present by day 100).  Unlike PTF11iqb which shifts from having a prominent blue peak early on, to a red peak in H$\alpha$ after day 500, SN 2013L has a persistent blue asymmetry until our last spectra on day 1509. SN 1998S also maintains a blue peak well past $\sim$ 1000 days. Elsewhere in the optical spectra the similarities end. Unlike PTF11iqb and SN 1998S, SN 2013L does not show nebular emission lines in the spectra except for the Ca II IR triplet, supporting the presence of strong CSM interaction which masks the ejecta emission.

For comparison we have added other non-IIn CCSNe SN 2004dj, SN 2007od, and SN 2011ja which also show asymmetric and multi-peaked H$\alpha$, and have compared the profiles both at early ($\sim$ 100 days) and late ($\sim$ 1000 days) times. Obvious in the early-time comparison is the lower velocities of the blue component of the non-IIn objects.  SN 2004dj exhibited a blue peak at -1770 km s$^{-1}$  \citep{2011ApJ...732..109M, 2006MNRAS.369.1780V}, SN 2007od showed a blue peak at roughly -1500 km s$^{-1}$ \citep{2010ApJ...715..541A}, and SN 2011ja at -1400 km s$^{-1}$ \citep{2016MNRAS.457.3241A}. The IIn SNe however have blue peaks residing between -3500 km s$^{-1}$ and -5000 km s$^{-1}$. There is a -5000 km s$^{-1}$ peak that begins to emerge at late times in SN 2007od, likely indicative of CSM/ejecta interaction much further from the explosion, but it is a separate component from the early peak at -1500 km s$^{-1}$.    
 
The measured CSM velocities for PTF11iqb, SN 1998S and SN 2013L are similar and indicative of a high mass extended progenitor, with 1998S exhibiting 40 km s$^{-1}$ speeds \citep{2015ApJ...806..213S}, PTF11iqb has a resolution limited CSM velocity of  $<$ 80 km s$^{-1}$ \citep{2015MNRAS.449.1876S} and SN 2013L at 84  km s$^{-1}$. The mass-loss rates are also similar, with an $\dot{M}$ $\sim$ 0.3--7.6 $\times$ 10$^{-3}$ M$_{\sun}$ yr$^{-1}$ of SN 2013L falling comfortably with both PTF11iqb ($\dot{M}$ $\sim$ 1 $\times$ 10$^{-4}$ M$_{\sun}$ yr$^{-1}$  \citep{2015MNRAS.449.1876S}) and SN 1998S ($\dot{M}$ $\sim$ 2 $\times$ 10$^{-5}$ $_{\sun}$ yr$^{-1}$  \citep{2001MNRAS.325..907F}), albeit on the higher side.  While both PTF11iqb and SN 1998S have mass-loss rates and progenitor wind speeds that are suggestive of extreme RSG, YSG, or YHG progenitors  the slightly elevated wind speed and $\dot{M}$ of SN 2013L may be more likely due to a YHG or LBV progenitor.

\subsubsection{SN 2005ip}
It is also worth taking a moment to discuss how SN 2013L is like and unlike the well studied IIn SN 2005ip.  While the H$\alpha$ luminosity of both objects began roughly equally, SN 2005ip is about 5 $\times$ more luminous (Figure \ref{fig:halum}) than SN 2013L at day  $\sim$ 730.  The R-band light-curve behaves similarly,  but with the maximum absolute magnitude of SN 2013L reaching roughly a magnitude brighter (Figure \ref {fig:LC}) before fading below that of SN 2005ip after 500 days.  Additionally the colour evolution over the first 100 days is quite different, with SN 2005ip staying much bluer at a constant (B--V) = 0.40, while SN 2013L was gradually becoming redder (Figure \ref{fig:colorev}).  Nebular and coronal lines were prominent in SN 2005ip, but not seen in SN 2013L. These differences could be attributed to asymmetries, as SN 2005ip seems to be mostly spherical unlike the disc-like shape inferred for SN 2013L. This could  account for the much brighter H$\alpha$ luminosity of SN 2005ip which has a much larger area of CSM that can interact with the SN ejecta.  

\section{Conclusions}
SN 2013L is an enduring SN IIn with long-lasting CSM interaction and prominent multi-peaked H$\alpha$ emission.  The multiple components arise from the geometry of the system and the mass-loss history of the SN progenitor.  Narrow emission comes from the ionization of the slow-moving wind from the progenitor, intermediate-width components from the ejecta/CSM interaction, and underlying broad emission from the freely expanding ejecta.  The blue-shifted hydrogen component that is seen in all of our epochs, and the lack of a red-shifted component, indicates that either the CSM material is distributed asymmetrically between us and the SN or that there is a disc or torus of CSM.  The narrow absorption which persists until the last observation on day 1509 suggests that we are seeing the system from roughly edge on.  The onset of CSM interaction by day 5 and persisting until our last observation on day 1509 suggests a dense and/or very extended CSM. If this object behaves anything like SNe 2005ip and 1988Z we may continue seeing signs of CSM interaction for years to come.

The narrow P-Cygni absorption indicates an outflow wind velocity between 80--130 km s$^{-1}$, speeds commonly associated with a YHG or quiescent LBV wind \citep{2014ARA&A..52..487S}. The observations we present here suggest that SN 2013L has a more asymmetric and less inclined CSM  than the similar IIn SNe 1998S or PTF11iqb, and were it to have a spherically distributed CSM would closely resemble SN 2005ip.  While the absolute magnitude of SN 2013L does not put it in the super luminous SN regime, it is bright enough to require an extreme amount of mass-loss in the years prior to explosion.   With objects such as SN 2013L and PTF11iqb, it is becoming more apparent that there are multiple progenitor scenarios that can likely give rise to early and strong CSM interaction, including RSGs in a binary system. The addition of CSM asymmetry in Type IIn SNe allows us to be able to create a continuum among the non-super luminous IIn, and opens the door for RSGs to be common progenitors for all but the brightest SNe IIn.

\smallskip\smallskip\smallskip\smallskip
\noindent {\bf ACKNOWLEDGMENTS}
\smallskip
\footnotesize
Special thanks to Melissa Graham for observations and data reduction of SN 2013L taken with LCO and to the anonymous referee for constructive comments and suggestions. This research has made use of the services of the ESO Science Archive Facility and  includes data gathered with the 6.5 meter Magellan Telescopes located at Las Campanas Observatory, Chile. This work makes use of observations from the LCO network, and is based in part on observations from Spitzer Space Telescope. Support was provided by the National Science Foundation (NSF) through grants AST-1210599 and AST-1312221 to the University of Arizona. This research has made use of the NASA/IPAC Extragalactic Database (NED) which is operated by the Jet Propulsion Laboratory, California Institute of Technology, under contract with the National Aeronautics and Space Administration.

\bibliographystyle{mnras}
\bibliography{SN2013L}
\bsp	
\label{lastpage}
\end{document}